\documentclass[conference,compsoc]{IEEEtran}
\usepackage{tikz}
\usepackage{amsmath}

\usepackage{comment}
\usepackage{spverbatim}
\usepackage{listings}
\usepackage{mdframed}
\usepackage{fancyvrb}
\usepackage{multirow}
\usepackage{multicol} 
\usepackage{tcolorbox}
\usepackage{colortbl}  
\usepackage{pifont}
\usepackage{makecell}
\usepackage{graphicx}
\usepackage{caption}
\usepackage{subcaption}
\usepackage{booktabs}
\usepackage{alltt}
\usepackage{rotating}
\usepackage{array}
\usepackage{multirow}
\usepackage{amsfonts}
\usepackage{amsmath}

\usepackage{url}

\definecolor{navyblue}{rgb}{0.0, 0.0, 0.5}

\newcommand{\new}[1]{{\color{black} #1}}

\newcommand{\para}[1]{{\vspace{4pt} \bf 
\noindent #1 \hspace{8pt}}}

\newcommand{\name}{{\tt Revelio}}

\newenvironment{packed_itemize}{
    \begin{list}{\labelitemi}{\leftmargin=1.2em}
    \setlength{\itemsep}{4pt}
    \setlength{\parskip}{0pt}
    \setlength{\parsep}{0pt}
    \setlength{\headsep}{0pt}
    \setlength{\topskip}{0pt}
    \setlength{\topmargin}{0pt}
    \setlength{\topsep}{0pt}
    \setlength{\partopsep}{4pt}
}{\end{list}}

\ifCLASSOPTIONcompsoc
  \usepackage[nocompress]{cite}
\else
  \usepackage{cite}
\fi
\ifCLASSINFOpdf
\else
\fi
\begin{document}

%

\title{Restoring Gaussian Blurred Face Images for Deanonymization Attacks}

\author{\IEEEauthorblockN{Haoyu Zhai\IEEEauthorrefmark{1}, Shuo Wang\IEEEauthorrefmark{1}, Pirouz Naghavi, Qingying Hao, Gang Wang
    \IEEEauthorblockA{University of Illinois Urbana-Champaign
    \\\{zhai11, shuow6, naghavi2, qhao2, gangw\}@illinois.edu}
}}
\maketitle

\def\thefootnote{\IEEEauthorrefmark{1}}\footnotetext{These authors contributed equally to this work.}
\begin{abstract}
Gaussian blur is widely used to blur human faces in sensitive photos before the photos are posted on the Internet. However, it is unclear to what extent the blurred faces can be restored and used to re-identify the person, especially under a {\em high-blurring} setting. In this paper, we explore this question by developing a deblurring method called \name{}. The key intuition is to leverage a generative model's memorization effect and approximate the inverse function of Gaussian blur for face restoration. Compared with existing methods, we design the deblurring process to be {\em identity-preserving}. It uses a conditional Diffusion model for preliminary face restoration and then uses an identity retrieval model to retrieve related images to further enhance fidelity. We evaluate \name{} with large public face image datasets and show that it can effectively restore blurred faces, especially under a high-blurring setting. It has a re-identification accuracy of 95.9\%, outperforming existing solutions. The result suggests that Gaussian blur should not be used for face anonymization purposes. We also demonstrate the robustness of this method against mismatched Gaussian kernel sizes and functions, and test preliminary countermeasures and adaptive attacks to inspire future work. 
\end{abstract}


%
\IEEEpeerreviewmaketitle

\section{Introduction}
\label{sec:intro}

With the rise of online social networks, search engines, and content-sharing platforms, billions of photos are circulating on the Internet, many of which contain {\em identifiable human faces}. For privacy considerations, users often {\em blur} the faces in a sensitive photo before posting it on the Internet. For example, news media may publish photos of crime scenes and blur the faces of the victims/offenders to protect their privacy~\cite{faceblur1, faceblur2}.
Similarly, people who post photos of civil unrest usually blur the faces of protesters~\cite{faceblur3, faceblur4}. 
Social media users who post photos of themselves/friends/strangers may choose to blur the faces if the photos capture sensitive, unflattering, or even illegal activities (e.g., drinking, stealing, or using drugs). 

Gaussian blur~\cite{HUMMEL198766} is among the most commonly used blurring algorithms~\cite{bethany2023towards, hill2016effectiveness} to smooth and remove high-frequency details in face images. This is done
by weighted-averaging on every pixel with its neighboring pixels where the weights are based on a Gaussian distribution (realized by a convolution kernel). Under a high-blurring setting, Gaussian blur can make the face unidentifiable to human eyes and thus is often used as an anonymization tool. The wide use of Gaussian blur, especially by lay users, is largely due to its availability in everyday photo-processing software and apps such as Adobe Photoshop~\cite{faceblurapps4}, Apple's Motion~\cite{motion}, Android's Blur Photo Editor~\cite{bpeditor}, PicsArt~\cite{picsart}, and ASPOSE~\cite{aspose}.

\para{Motivation.} 
In this paper, we ask one basic question: {\em To what extent can adversaries restore a Gaussian blurred face, and identify the person in the photo}. We particularly explore this question under a {\em high-blurring} level to push the limit of the deblurring method. Prior works have explored related questions but under different deanonymization contexts. For example, Cavedon et al.~\cite{cavedon2011getting} showed that {\em image pixelization} could be reversed using a Maximum A Posteriori (MAP) method but their method cannot be applied to Gaussian blur. In addition, their method is customized for video streams (i.e., requiring multiple video frames) instead of a single photo. Hill et al.~\cite{hill2016effectiveness} used a Hidden Markov Model (HMM) to restore {\em redacted text} on documents. This method relies on the fixed set of English characters and digits for text recovery, but does not apply to human faces.

\begin{figure}[t]
  \centering
  \includegraphics[width=0.9\linewidth]{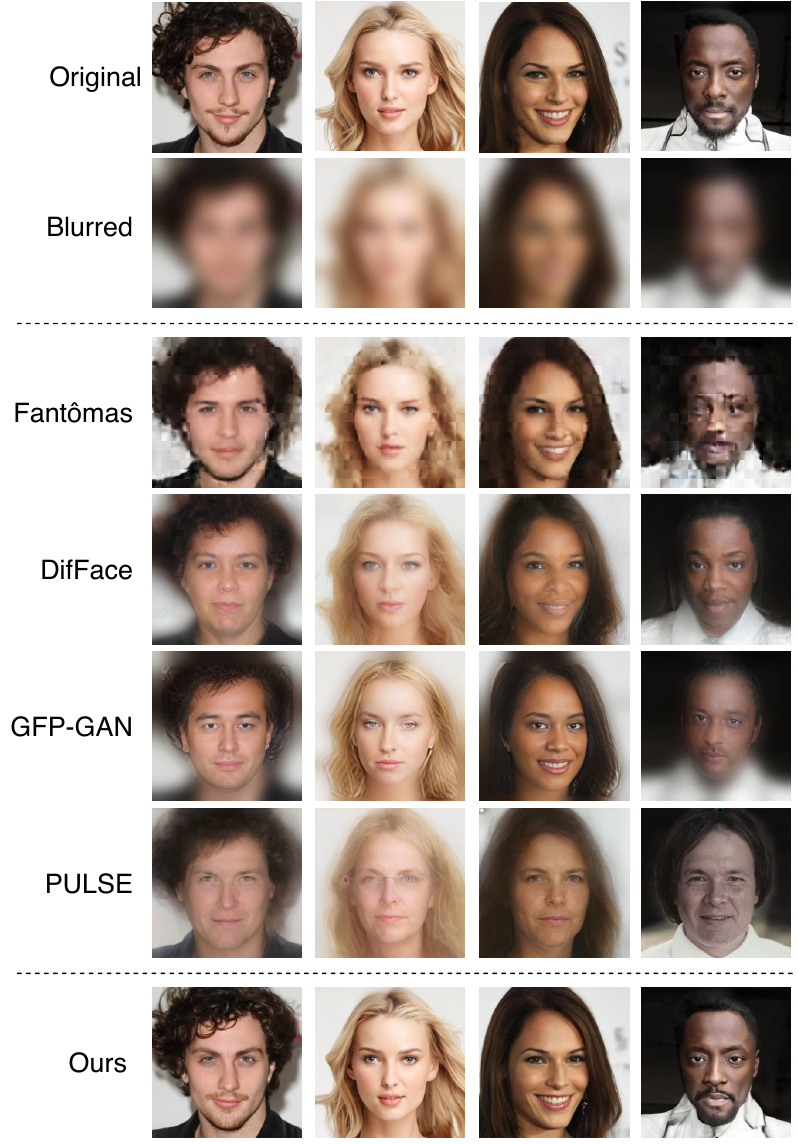}
  \caption{The original face has been blurred with Gaussian blur (kernel size $K=81$) to hide the user identity. Our proposed method (\name) can restore the blurred face with a higher fidelity with respect to the original identity. 
  }
  \label{fig:example}
  \vspace{-0.13in}
\end{figure}

More recently, related work from the machine learning community seeks to address the (blind) image restoration problem~\cite{gfpgan,difFace, xia2023diffir,pulse,todt2024fantomas}. Their goal is to restore high-quality images from degraded photos without prior knowledge of the degradation process. 
These works {\em cannot answer our research questions} for two reasons. First, most existing solutions are designed to restore unintentional image blur caused by camera shakes, poor lighting conditions, and low-quality cameras.  
The target degradation strength is usually low. In contrast, we focus on Gaussian blur {\em intentionally applied} for privacy protection, and thus the expected blurring level is much higher. As shown in Figure~\ref{fig:example}, under a heavy-blur setting (with a Gaussian kernel size of 81), existing solutions have a subpar performance. Second and more importantly, existing solutions are not designed to be {\em identity-aware}. They focus on re-generating facial details but the face does not need to be that of the original person/identity (Figure~\ref{fig:example}). In contrast, we want to preserve the identity of the original image. Note that, one recent solution Fantômas~\cite{todt2024fantomas} indeed targeted face anonymization scenarios. However, they also did not explore restoring faces from highly blurred images (their kernel size is only 29). \new{We re-trained their method under a heavy-blur setting, and Figure~\ref{fig:example} shows that the face restoration is still suboptimal. }


\para{Our Approach.}
To fill in the gap and answer our research questions, we develop a system called \name{}. The key idea is to (1) use a conditional Diffusion model to approximate the inverse of the Gaussian blur function, and (2) use a large reference database and an identity retrieval model to augment the face restoration process. 
The intuition is two-fold. First, we utilize the memorization effect~\cite{somepalli2023understanding} in diffusion models. We assume the adversary can construct a large {\em reference database} $D$ of face images. This is a realistic assumption given the aggressive image collection from social media services and search engines, which can also be done by other parties via web crawling. By training the diffusion model with such a reference database, we use the memorization effect to synthesize face details based on the matching identity. \new{Importantly, even if the target identity is {\em not} in the reference database $D$, it is still possible for the model to perform face restoration with {\em similar-looking} identities in the reference database  (validated in \S\ref{sec:open-world}).
}
Second, while it is impossible to construct a lossless inverse function for Gaussian blur, we modify the diffusion model to {\em approximate} the mapping from the blurred face to the original clear face. This is done by making the denoising process {\em conditioned} on the blurred input to preserve the original identity. 

Based on these intuitions, \name{} takes a Gaussian blurred image $x$ as input, and first performs a {\em preliminary restoration} to produce an image $y_b$ using a base diffusion Model. Then it uses $y_b$ to query the reference database to find a potentially matching identity. If the blurred image contains a known identity, the system retrieves reference images of this person and uses these images to {\em fine-tune} the base model to enhance the fidelity of face restoration. However, if the person is never seen (i.e., not in the reference database), \name{} can detect this is an out-of-distribution (OOD) identity, but still uses other ``similar-looking'' faces to recover the face details.

\para{Evaluation.}
We evaluate \name{} using public face image datasets CelebA-HQ~\cite{karras2017progressive} and \new {FFHQ~\cite{karras2019stylebasedgeneratorarchitecturegenerative}. 
}
We construct a reference database of 28,000 images and perform testing on images that {\em never appear} in the reference database. We test both light-blur (kernel size 37) and heavy-blur (kernel size 81) settings, and compare \name{} with existing face restoration solutions (GFP-GAN~\cite{gfpgan}, DifFace~\cite{difFace}, PULSE~\cite{pulse}, and Fantômas~\cite{todt2024fantomas}), and a Parrot Recognition method~\cite{parrot_recog}. We have three main observations. 
First, \name{} outperforms existing methods with respect to the restored image quality, and more importantly, the {\em fidelity to the original identity}. The advantage of \name{} is more significant under the heavy-blur setting. 
Second, adversaries can use \name{} to successfully retrieve the identity of the blurred image when the identity is in the reference database. Under a heavy-blur setting, the identity retrieval accuracy achieves 95.9\%. Importantly, even when the identity is {\em never seen} (OOD identity), \name{} can still use other similar-looking faces to restore the face (with minimal quality degradation). 
Third, \name{} exhibits robustness against images blurred with unknown Gaussian kernel sizes. Its built-in kernel size estimator can accurately infer the kernel size with a mean absolute error (MAE) lower than 1, and can tolerate kernel size mismatches within an offset of 6.     

While defense is not the main focus of the paper, we have experimented with a few countermeasure ideas to disrupt the adversaries' models. We show that, \new{while some of the countermeasures (e.g., image compression) can affect the vanilla attack, the impact can be largely mitigated by {\em adaptive attacks}.} Further research is still needed to develop more secure and robust face anonymization methods. \new{We have responsibly disclosed our findings to related parties including  OpenCV, PyTorch, Adobe Photoshop, Apple (iOS SDK team) and FTC (Federal Trade Commission). Further ethics discussions are in \S\ref{sec:ethics}.
}





\para{Contributions.} Our paper has three key contributions:  

\begin{packed_itemize}
\item {\bf New method}: we developed an {\em identity-aware} deblurring method to restore Gaussian blurred face images. It combines a conditional Diffusion model with an identity retrieval method to restore the blurred face images while preserving identity fidelity.     

\item {\bf New evaluation}: unlike prior works, we specifically target a {\em high-blurring} level to test the deblurring algorithm, which is more aligned with the face anonymization threat model. We show that our method outperforms existing solutions under such settings.  

\item {\bf New tools}: we plan to responsibly share our code with other researchers to facilitate the development of defense methods. We have discussed our considerations regarding how to prevent potential abusive usage of the code 


\end{packed_itemize}






\section{Background and Related Work}
\label{sec:back}

\subsection{Gaussian Blur}
\label{sec:gaussian}

Gaussian blur~\cite{HUMMEL198766} is a widely used blurring method available in many popular photo-processing software such as Adobe Photoshop~\cite{faceblurapps4}, Apple's Motion~\cite{motion}, Blur Photo Editor~\cite{bpeditor}, PicsArt~\cite{picsart}, and ASPOSE~\cite{aspose}. It is also supported by mobile  SDK~\cite{sdk1,sdk2}.
Gaussian blur removes high-frequency details from the image by convolving the image with a Gaussian function. It superimposes a 2D Gaussian distribution over a group of pixels in the image and computes new RGB values by weighted averaging every pixel with its neighboring pixels. The weights are based on the Gaussian distribution. Below shows the Gaussian function in two dimensions:
\begin{equation}
    G(i,j) = \frac{1}{2\pi \sigma^2}e^{-\frac{i^2 + j^2}{2\sigma^2}}
\end{equation}
This function creates the convolution {\em kernel} applied to every pixel in the original image. Here, $i$ and $j$ specify the location coordinates from the center pixel (0, 0). We use $K$ to represent the convolution kernel size, which also defines the range of $i$ and $j$. For example, if the kernel size $K = 3$, then $i$ and $j$ would range from -1 to 1 (inclusive). In this paper, we use a Gaussian blur with a square kernel. $K$ needs to be a positive {\em odd integer} such that there is one center pixel in the kernel. Note that the Gaussian distribution is parameterized by $\sigma$ (standard deviation), which also influences the strength of the blurring. A larger $\sigma$ leads to a flatter/wider Gaussian distribution, 
leading to a stronger blurring effect. 
In a typical Gaussian blur implementation, $\sigma$ needs to be scaled in proportion to the kernel size $K$. Otherwise, if we have a large kernel radius but a small $\sigma$, the far-away pixels would have little impact despite the large kernel size. As a result, real-world Gaussian blur implementations often use a fixed function to compute $\sigma$ based on the user-specified kernel size. For example, both PyTorch~\cite{paszke2017automatic} and OpenCV~\cite{opencv_library} implemented Gaussian blur using this dependency function: $\sigma = 0.3 * ((K - 1) * 0.5 - 1) + 0.8$.

\subsection{Blind Image Restoration}
\label{sec:blindface}
The machine learning community has worked on {\em blind image restoration}, which is related to but is different from our problem. Blind image restoration aims to recover high-quality images from degraded images without prior knowledge of the degradation.
There are three main categories of methods. First, researchers have used Generative Adversarial Networks (GANs) for image restoration such as GFP-GAN~\cite{gfpgan}, GPEN~\cite{gpen} and PSFR-GAN~\cite{psfrgan}. Second, researchers also use diffusion models for blind image restoration~\cite{ lin2024diffbir,saharia2022palette,xia2023diffir,wang2023zeroshot,zhu2023denoising}, some of which can be used on face images (e.g., DifFace~\cite{difFace}). 
However, they often focus on colorization, inpainting, uncropping, and JPEG restoration, but not Gaussian blur restoration for human faces. 
Third, super-resolution models~\cite{saharia2021SR, pulse, SRDiff} transform low-resolution images into high-resolution ones (e.g.,  PULSE~\cite{pulse}). 



There are two key differences between the existing work and this paper. First, {\em application scenarios}. Most of these solutions are designed to handle unintentional image blur (e.g., caused by camera shakes), and their expected degradation strength is not very high. In contrast, we focus on {\em intentional} image blur for privacy protection. The expected blurring level needs to be high to hide facial identity. Second, {\em identity-awareness}. Most solutions are not designed to be identity-aware. The restored face details may not be those of the same person (Figure~\ref{fig:example}).

\subsection{Image Restoration vs. Privacy Protection} 
\label{sec:privacy-lit}

Prior works from the security community have looked into how to restore images that have been blurred or pixelated for privacy/anonymization purposes. 
For example, Cavedon et al.~\cite{cavedon2011getting} show that {\em pixelated} videos can be recovered using a Maximum A Posteriori (MAP) method. However, their inverse function is specifically designed for pixelization rather than Gaussian blur. They also require {\em multiple frames} in a video stream for the restoration. Hill et al.~\cite{hill2016effectiveness} introduce an attack method to recover the original text in redacted documents (from pixelization and blurring), but it does not handle human face images.


More recently, researchers have used machine learning methods to restore face images under anonymization. For example, Todt et al.~\cite{todt2024fantomas}  evaluated 15 image anonymization techniques (including Gaussian blur) and introduced Fantômas, an auto-encoder-based image restoration method. Unfortunately, this work did not perform face restoration under a {\em high degradation setting}.
For Gaussian blur, they only used a small kernel size ($K=29$). \new{In \S\ref{sec:eval}, we show that Fantômas cannot effectively handle heavily blurred face images ($K=81$) even after re-training.} 

\new{Other related works proposed ``Parrot Recognition'' to directly re-identify anonymized images {\em without} performing image restoration~\cite{mcpherson2016defeating,parrot_recog}. 
The idea is to train the classifier on blurred images and then perform classification on blurred images to recognize their identity. While the classification accuracy is higher than random guessing, it is still challenging for an accurate reidentification under a heavy-blur setting when there are a large number of candidate identities (e.g., over 1000). The accuracy is around 65\% as shown in a recent work~\cite{parrot_recog}. Considering that parrot recognition does not restore blurred faces, it only addresses part of our threat model (see \S\ref{sec:threat-model} for details).
}

\subsection{Facial Recognition and Attacks}
\label{sec:face}
There is a large body of related work on facial recognition systems~\cite{deng2019arcface, deepface, wang2018cosface, liu2017sphereface, schroff2015facenet, sun2020circle, meng2021magface, kim2022adaface}, and adversarial attacks against them~\cite{wu2024uniid, garofalo2018fishy, sharif2016accessorize, li2014understanding}. 
In addition, researchers have also re-purposed some of the attack methods as privacy protection tools~\cite{shan2020fawkes, jin2024faceobfuscator,chen2023face, chow2024diversity}. This line of work is different from ours because they operate on {\em clear} face images rather than Gaussian blurred images. Also, these privacy protection mechanisms have very different goals. They seek to fool facial recognition models, but the human faces in the images should still be visible and recognizable to human eyes.

\section{Threat Model}
\label{sec:threat-model}










We focus on photos that contain identifiable human faces. To protect user privacy, Gaussian blur is used to blur the {\em whole face} before the photo is posted on the Internet. 



We assume the attacker has access to the blurred version of the photo (denoted as $x$), {\em but does not have access to} the original photo before blurring (denoted as $y$). The attacker's goal is to (1) restore the blurred photo such that the recovered face looks similar to the original face (i.e., face restoration), and (2) reveal the identity of the blurred face by matching it to a set of known identities (i.e., deanonymization). \new{We believe that {\em deanonymization} and {\em face restoration} are two different levels of privacy violations, and achieving both presents a stronger attack.}  

In this case, we assume the attacker maintains a large {\em reference database} of human face images (denoted as $D$). The availability of such a dataset is a realistic assumption given the aggressive data collection of user images by social media services, search engines, AI companies, and even individuals via web crawling. The attacker only focuses on the face area---they can crop the blurred face region (e.g., in a square) before running the deblurring attack. 


In our study, we consider two possible scenarios. First, in-distribution attack (or closed-world attack): the attacker's reference database $D$ contains this victim's {\em other photos} (not $y$). The attacker thus aims to restore the face and link the known identity to the blurred image, \new{achieving both deanonymization and face restoration}.  
Second, out-of-distribution (OOD) attack (or open-world attack): the attacker's reference database $D$ never indexes any photos of this victim. In this case, the attacker should be able to tell that the victim is not in the reference database. However, the attacker can still achieve face restoration, i.e., recovering the blurred face to look similar to the victim's original face (using other face images in $D$). \new{While OOD attack cannot achieve {\em immediate} deanonymization, it still has practical implications through {\em face restoration}. For example, by publishing the ``deblurred'' photo, the person in the photo could be potentially re-identified in an ad-hoc manner by the viewers who know the person in real life (e.g., friends/families/supervisors). Also, if the target is a wanted criminal, law enforcement could use the restored photo (as a police sketch) for their investigation. 
}

\begin{figure*}[t]
  \centering
  \includegraphics[width=\linewidth]{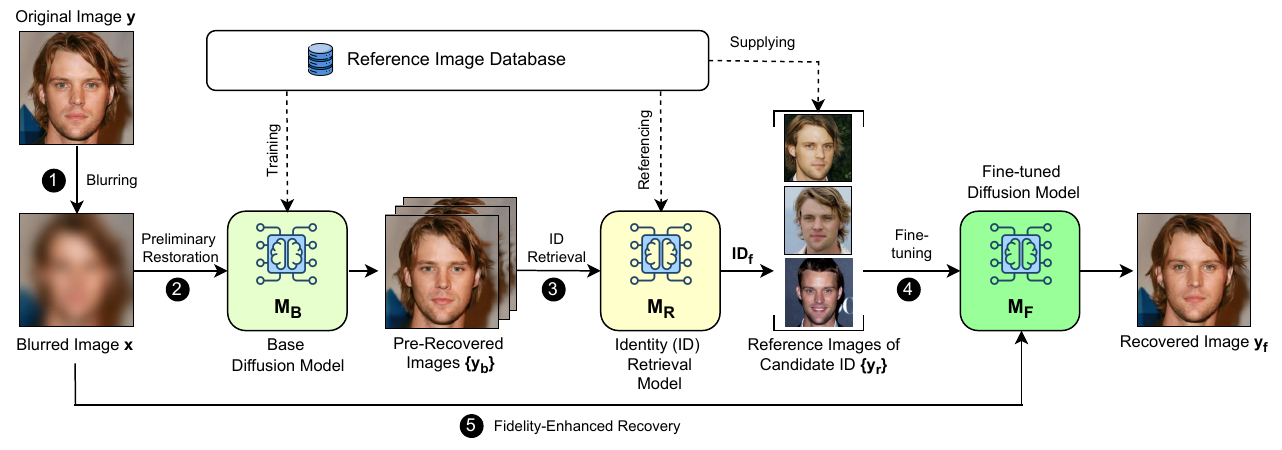}
  \caption{System workflow. 
The original clear image $y$ is Gaussian blurred to generate image $x$ (\ding{182}). This blurred image $x$ is the input to \name{}. 
To recover the original face, we first use a Base Model $M_B$ to perform preliminary restoration (\ding{183}). Then in step \ding{184}, we use pre-recovered image $y_b$ for identity retrieval. In step \ding{185}, we use the matched identity to find more reference images to fine-tune the model. Finally, using the fine-tuned model ($M_F$), we can further improve the fidelity of the restored image (\ding{186}).
  }
  \label{fig:pipeline}
  \vspace{-0.12in}
\end{figure*}

\section{Methodology}
\label{sec:method}

We develop a system called \name{} to restore blurred face images and recover their original identities. The system design of \name{} is shown in Figure~\ref{fig:pipeline}. 



\subsection{Intuitions and Design Overview}
\label{sec:intuition}


\para{Challenges.} First, Gaussian blur introduces significant ``information loss'' to an image as introduced in \S\ref{sec:gaussian}. 
Existing works such as DDRM~\cite{kawar2022denoising} and SNIPS~\cite{kawar2021snipssolvingnoisyinverse} treat the general image restoration problem as a {\em linear inverse} problem; however, the degradation matrix of Gaussian blur is ill-conditioned for this formulation due to its irreversible information loss~\cite{HUMMEL198766}. 
ML-based restoration methods~\cite{gfpgan,difFace} are not designed for intentionally applied heavy blurring for privacy reasons (\S\ref{sec:blindface}).


\para{Design Intuitions.} We design \name{} based on two key intuitions. First, we utilize the memorization effect of generative models. Recent research~\cite{somepalli2023understanding, extractingtrainingdata} shows that generated images from diffusion models are partially replicating the training data. In our scenario, this is not necessarily a weakness but an opportunity to construct attacks. Even though the model is {\em not} trained on the clear version of the input image, as long as the model has seen this person's other images, the memorization of training data could lead the model to reproduce this person's facial features.  Second, while it is impossible to construct a lossless inverse function for Gaussian blur, we can use a diffusion model to {\em approximate} the inverse function. The information loss can be {\em partially} recovered by learning the features of general human faces, especially faces that look similar to the target. 



\subsection{Face Restoration}
\label{sec:fr}
\label{sec:ke}


We start by constructing the Base Model $M_B$ for face restoration (step \ding{183} in Figure~\ref{fig:pipeline}). Given a blurred image $x$, $M_B$ aims to restore it to a clear image that looks similar to the original image $y$. 





\para{Training: Overview.}
We train $M_B$ using the {\em reference database} $D$ owned by the adversary. Using $D$, we first construct the training dataset by applying Gaussian blur on the images to produce ``ground-truth'' pairs of blurred images $x$ and their originals $y$ as tuples: $\{(x, y)\}$. Using this dataset, the model learns the inverse mapping from the blurred image $x$ to the corresponding clear image $y$. The idea is to approximate the conditional distribution $p(y|x)$ (which represents the inverse of Gaussian blur). Instead of using the traditional Denoising Diffusion Probabilistic Model (DDPM)~\cite{ho2020DDPM} (which only captures the distribution $p(y)$ of generic face images), we use conditional diffusion~\cite{saharia2022palette} such that the generated faces are faithful to the original face's identity (i.e., conditioned by the input $x$). 


\para{Diffusion Model's Forward Process.}
A diffusion model has a forward process and a reverse (denoising) process. The forward process gradually adds Gaussian noise (not Gaussian blur) to ``destroy'' the clear image until the image turns into pure Gaussian noise. Then, we learn to recover the original image by modeling the reverse process. 
For our forward process, we start by drawing an image pair $(x,y)$ from the training dataset. We set $y_0 = y$, and then a forward Markov chain gradually adds Gaussian noise into $y_0$ with a total of $T$ steps: 

\begin{equation}
q(y_t | y_{t-1}) = \mathcal{N}(y_t | \sqrt{\alpha_t} y_{t-1}, (1-\alpha_t)\mathbf{I})
\end{equation}
Here $t$ represents a time step from 1 to $T$. $\mathcal{N}$ and $\mathbf{I}$ denote a Gaussian distribution and the identity matrix, respectively. $\alpha_t$ is a hyper-parameter between 0 and 1 which defines the variance of the Gaussian noise added in each step. We can rewrite the forward process into a closed form. Given the initial clear image $y_0$ and $y_t$ at any step, we can represent the distribution $y_{t-1}$ as the following: 

\begin{equation}
    q(y_{t-1} | y_0, y_t) = \mathcal{N}(y_{t-1} | \mu, \sigma^2\mathbf{I})
\label{eq:firward-noise}
\end{equation}
where $\mu = \frac{\sqrt{\gamma_{t-1}}(1-\alpha_t)}{1-\gamma_t}y_0 + \frac{\sqrt{\alpha_t}(1-\gamma_{t-1})}{1-\gamma_t}y_t, \sigma^2 = \frac{(1-\gamma_{t-1})(1-\alpha_t)}{1-\gamma_{t}}, \text{and } \gamma_t = \Pi_{s=1}^t \alpha_s$. Note that $\gamma_t$ can represent the strength of the current noise level, which will be used in the denoising process later. $\mu$ and $\sigma$ here are parameters of Gaussian noise (not Gaussian blur). 




\para{Conditional Denoising Process.}
For the de-noising process, we learn to recover the clear image $y$ by optimizing a neural network model $f_{\theta}$ that is {\em conditioned} on the blurred image $x$. This model $f_\theta$ takes as inputs the blurred image $x$, a noisy image $y_t$ at any step, and its current noise level $\gamma_t$, and predicts the noise vector $\epsilon$. The noise vector $\epsilon$ will be used to denoise $y_t$ to restore the previous version $y_{t-1}$.
Here we use the same objective function for training the network $f_\theta$ used by~\cite{saharia2022palette}: 
\begin{equation}
    \mathbb{E}_{(x,y)}\mathbb{E}_{\epsilon, \gamma} \left\Vert f_{\theta}(x, y_t,  \gamma_t) - \epsilon \right\Vert_2^2
\label{eq:lossfunc}
\end{equation}


Optimizing Eqn.~\ref{eq:lossfunc} is the key to face restoration. Here we use a U-Net architecture for $f_\theta$, {\em concatenating the blurred image $x$ with $y_t$ as the input}. In this way, the blurred image $x$ serves as a prior that contains information about what the clear image should look like. Without $x$, the model would denoise $y_t$ blindly, leading to the loss of {\em identity} information (i.e., only generating generic human faces). By conditioning the denoising process on the blurred image $x$, the model learns to recover the clear image $y$ while using the blurred $x$ as guidance, preserving the identity of $x$. This denoising process learns to remove both the Gaussian noise and the Gaussian blur effect step by step. We also use the attention mechanism on the U-Net architecture. The U-Net architecture captures image features and expands them through multiple layers. At the bottleneck layer, the network encodes localized facial features (e.g., the eyes and the mouth). We apply an attention mechanism at this layer, encouraging the model to restore fine-grained face details. 



\para{Inference.} After the above training process, we obtain a $f_\theta$ that can approximate $y_{t-1}$ given $x, y_t, \text{and } \gamma_t$. Specifically, Eqn.~\ref{eq:firward-noise} now can be rewritten as the following to calculate $y_{t-1}$: 


\begin{equation}
    y_{t-1} \leftarrow \frac{1}{\sqrt{\alpha_t}}\left( y_t - \frac{1-\alpha_t}{\sqrt{1-\gamma_t}}f_\theta(x,y_t, \gamma_t)\right) + \sqrt{1-\alpha_t}\epsilon_t
\end{equation}
Thus, during the inference time, given an input blurred image $x$ and a Gaussian noise vector $y_T$, we can restore and refine the image iteratively to recover facial details while preserving the identity of $x$. The model is trained on a large-scale face image dataset ($D$), and we expect the model to memorize high-fidelity details related to given identities. During the inference time, we expect the model to restore the facial details leveraging the memorization effect (e.g., using other images of $x$'s identity in the database, or faces of other similar-looking identities).

\para{Practical Consideration: Kernel Size Estimation.}
Note that the above model $M_B$ is trained under a specific kernel size for Gaussian blur. The kernel size $K$ is set when we construct the training data for $M_B$. 
In practice, adversaries may not know the kernel size used by the input image $x$. In this case,  adversaries can either (1) try different models (trained with different kernel sizes) and assess the recovered image quality, or (2) proactively estimate the kernel size based on input image $x$. We believe the second option is more cost-efficient. Below, we develop a kernel size estimator for this purpose. 


Given an input blurred image $x$, the goal of the model is to infer the kernel size $K$. To train the model, we first construct a {\em synthetic training dataset} with ``ground truth''. This is done by randomly sampling face images and applying Gaussian blur on each image using a series of kernel sizes. Then we train a kernel size estimator using a regression model.
Here, we use a pre-trained model EfficientNetV2 Large~\cite{tan2021efficientnetv2} as the pre-trained model to improve the model performance (pre-trained on ImageNet~\cite{deng2009imagenet}). Then we replace the classification layer with a single output layer for regression to estimate the kernel size $K$. We obtain the model size $K$ by rounding the regression estimation to the nearest odd integer value. Then the $K$ will be used to select the proper model $M_B$ for the face restoration of $x$.

\subsection{Identity Retrieval}
\label{sec:ir}

As shown in Figure~\ref{fig:pipeline} (\ding{184}), after using $M_B$ to perform preliminary face restoration, we expect the recovered face image $y_b$ to be somewhat similar to the original image but is not yet of high quality and fidelity. Our hypothesis is that image $y_b$ is {\em good enough} to recover the {\em identity} of the target person if this person is in the reference database.



\para{Face Embedding.}
To perform identity retrieval, the adversary first needs to map the face images in the reference database $D$ to an embedding space. 
In this embedding space, face images of the same person (identity) should be clustered together while face images from different identities should be mapped further away. To do so, we utilize an open-source face recognition system~\cite{face_recognition} to perform face image embedding. We select this model for its high facial recognition performance as it achieves an accuracy of 99.38\% on the LFW (``Labeled Face Images in the Wild'') benchmark~\cite{LFWTech}. 


\para{Identity Retrieval Function.}
A naive method is to compute the {\em average} distance between the input image $y_b$ and other images of existing identities in the embedding space to find the nearest neighbors for identity detection. However, this naive idea is easily affected by the {\em stochastic process} of face generation of the diffusion model $M_B$. 
When we run $M_B$ multiple times on the same input, the generated images have significant variance, especially when the blurring level is high. 
For certain rounds, the generated images are highly similar to the target person, while for other rounds the generated images look very different.
The intuitive explanation is that the {\em memorization effect}, when triggered, can produce face details of the original person. However, the memorization effect is not always triggered the same way (or triggered at all) each round. 

To this end, we determined that taking an {\em average distance} is not the best option. Instead, we run $M_B$ on the same input multiple times (i.e., $n$ rounds) to generate a set of $\{y_b\}$, and then rely on the {\em shortest distance} to determine the identity. The intuition is that within the multiple rounds of face generation, one or a few rounds will trigger a strong memorization effect to generate face images close to the true identity. We rely on these rounds (with the shortest distance) to maximize the success of identity retrieval.   
We take $n=50$ as our default settings. For each pre-recovered image in the set $\{y_b\}$, we calculate the distance between the restored image and its nearest neighbor in the reference database. We then take the identity (ID) of the nearest neighbor as the winning identity for this round. Finally, the most frequently appearing winning identity among the $n$ rounds would be selected as the final identity for input $x$, which is denoted as $ID_f$.  

\para{Detecting Out-of-Distribution (OOD) Identities. } 
The above identity retrieval process will make a mistake when the identity of the blurred image $x$ is never indexed in the reference database. In other words, the person in $x$ has no other images in the reference database and is never seen by the diffusion model. We call such identities as out-of-distribution (OOD) identities. To detect blurred images of OOD identities, we use a simple shortest-distance threshold. 
More specifically, out of the $n$ rounds of restoration, we identify the lowest distance between any of the pre-recovered images in $\{y_b\}$ and their nearest neighbors in the reference database. We denote this lowest distance as $ld_x$. If $ld_x$ is higher than a threshold $d$, we will determine that $x$ has an OOD identity. 
We will explore how the threshold affects OOD detection accuracy later in \S\ref{sec:open-world}.

\subsection{Fidelity Enhancement}
\label{sec:fi}

Once the identity has been recovered, we retrieve related images from the {\em reference database} to further improve the fidelity of image restoration. This step corresponds to steps \ding{185} and \ding{186}. 
The pre-recovered image $y_b$ may still carry random facial features from the stochastic diffusion process. To further improve fidelity, we fine-tune the base model. First, based on the retrieved identity $ID_f$, we obtain a set of reference images of this identity from the reference database (denoted as $\{y_r\}$). These images are of the same person but do not contain the target image $y$. Second, using these images as the fine-tuning dataset, we adapt the attention block fine-tuning technique~\cite{moon2022finetuning} to fine-tune $M_B$. We freeze all residual blocks in the U-Net backbone architecture in the base model except the attention blocks, which helps to preserve the identity information during fine-tuning.   
Finally, the fine-tuned model $M_F$ can generate image $y_f$, which is expected to be of higher fidelity to the ground-truth image. 


\section{Evaluation}
\label{sec:eval}

In this section, we evaluate the effectiveness of \name{} and compare it with existing solutions.

\begin{figure}[t]
  \centering
  \includegraphics[width=0.72\linewidth]{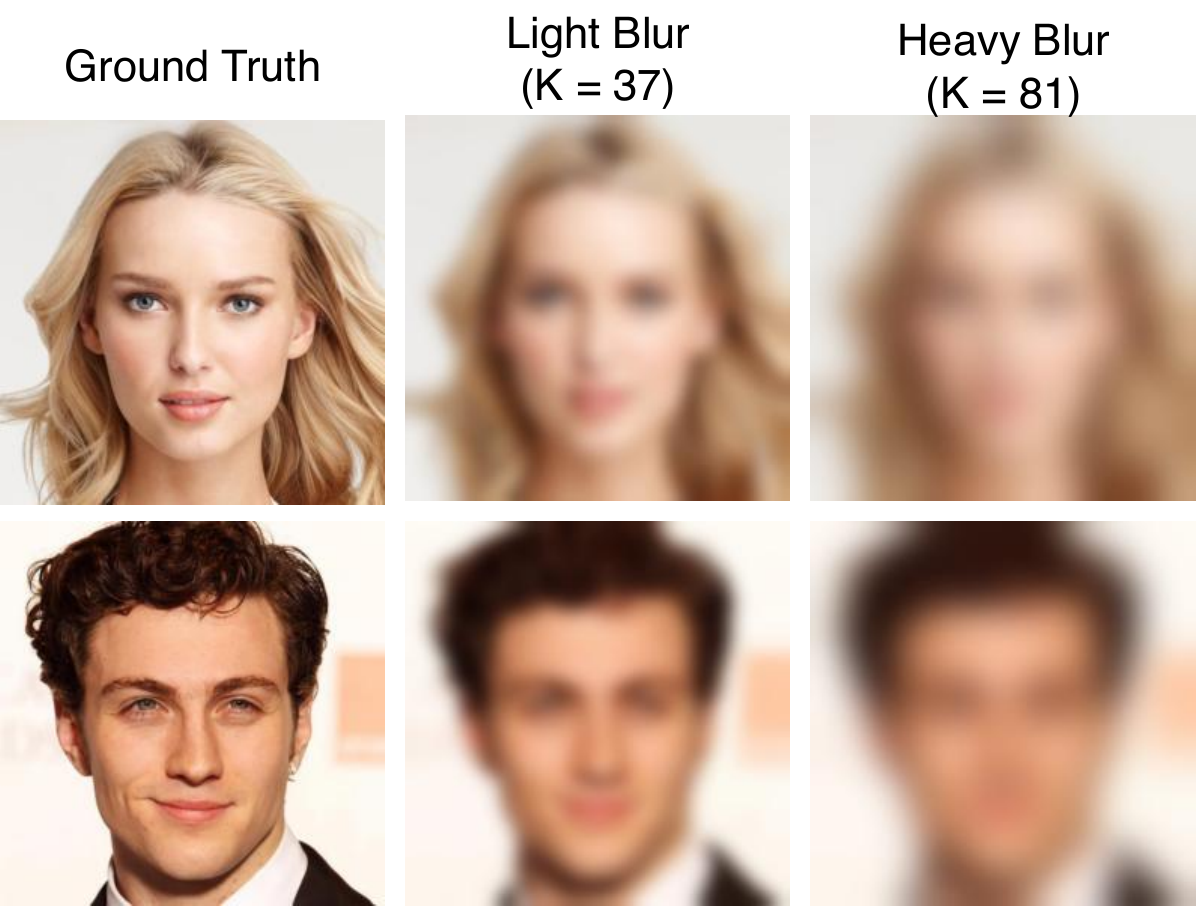}
  \caption{We use two different levels of Gaussian blur. The {\em light} blur setting has a kernel size $K = 37$. The {\em heavy} blur setting has a kernel size $K = 81$.}
  \label{fig:blur}
  \vspace{-0.12in}
\end{figure}

\subsection{Dataset and Setup}

\para{Gaussian Blur Setup.} We use PyTorch's implementation of Gaussian blur as the target for proof-of-concept. The same implementation is also used by OpenCV. 
We use kernel size $K$ to control blur levels (the standard deviation $\sigma$ of the Gaussian distribution is dependent on $K$).
We use two different blur levels including light blur and heavy blur (see Figure~\ref{fig:blur} for examples). Given the image size of $256 \times 256$ images, we apply Gaussian blur with kernel size $K=37$ for ``light blur'' and use kernel size $K=81$ for ``heavy blur''. $K=37$ roughly matches the degradation level of existing works. 
As discussed in \S\ref{sec:blindface}, existing works used Gaussian blur as a generic way to degrade image quality (rather than for privacy protection), and thus did not use high blurring levels. As shown in Figure~\ref{fig:blur}, under light blur, the blurred image does not fully hide the person's facial details. In contrast, under heavy blur, the images lose significantly more details, making it harder to re-identify the person. Our experiment will primarily focus on the high-blur setting given it is more challenging for attackers. In our experiments, if not otherwise stated, \name{} uses the matching kernel size to train the corresponding model. In \S\ref{sec:ks-exp}, we will evaluate the attacker's ability to predict the kernel size and its tolerance of mismatched kernel sizes.

\para{Datasets.}
Our experiment mainly uses the CelebA-HQ dataset~\cite{karras2017progressive}, which includes 30,000 images, and 6,217 labeled identities.
\new{Another dataset FFHQ~\cite{karras2019stylebasedgeneratorarchitecturegenerative} is used for OOD evaluation (\S\ref{sec:open-world}).
}
We selected these datasets because they have been widely used by existing research~\cite{gfpgan, gpen, psfrgan, difFace, kawar2022denoising, lin2024diffbir}, making it easy for result reproduction. Both datasets contain high-quality face images. Importantly, CelebA-HQ includes {\em identity labels} (rare among public datasets). Each person (identity) has multiple images (5 images per identity on average). 
\new{We did not use datasets such as MegaFace~\cite{kemelmacher2016megaface,nech2017level} and LFW~\cite{LFWTech}, either because the dataset was decommissioned or the image quality is low. }


We randomly split the images in CelebA-HQ into the training set (28,000 images) and the testing set (2,000 images). 
The training set contains 6,084 unique identities, and the testing set contains 1,521 unique identities (with 1,388 identities overlapping between the training and testing sets). Despite the overlap of identities, a specific image can only appear in either the training set or the testing set but not both. 
We train our Base Model $M_B$ with 28,000 training images for 1,000 epochs. Due to computing resource limitations, the dimensions of the images are set to $256\times256$ during training and testing. The {\em training set} (28,000 images) also serves as our reference database for identity retrieval.  

\para{Experiment Setup.}
Our experiment covers both closed-world (in-distribution attack) and open-world (OOD attack) settings. The main experiment uses the closed-world setting to assess the design choices in \name{}. Under this setting, the identities of the blurred images are {\em included} in the reference database. For open-world settings, we test identities that are {\em not included} in the reference database. 



For the closed-world setting, we randomly select 50 identities and a total of 97 images from the {\em testing set} to apply Gaussian blur. 
These identities have other images in the reference database (i.e., the training set) but these 97 testing images never appear in the training set.    
For the open-world setting, we manually construct a set of out-of-distribution (OOD) identities and ensure (1) the testing images never appear in the training set; and (2) these identities have no other images in the training set either. In other words, these identities are never seen by the model. We obtain 97 of such images from 91 OOD identities (see \S\ref{sec:open-world} for details).

\subsection{Evaluation Metrics and Baseline Methods}
\label{sec:metric}


\para{Evaluation Metrics.}
We use different metrics to evaluate the {\em identity retrieval} and {\em face restoration}.  
For identity retrieval (model $M_R$), we evaluate its performance using an {\em Identity Retrieval Accuracy} (IRA) which measures the ratio of blurred images for which the model correctly retrieves their identity out of all the testing images. 

For the face restoration (models $M_B$ and $M_F$), we use {\em standard} metrics to assess the fidelity of the restored images and the overall image quality. 
Fidelity evaluation measures how similar the restored image is in comparison with the original ground-truth image. First, we use PSNR~\cite{psnr} to make a similarity comparison at the pixel level for the two images. Second, we use LPIPS~\cite{lpips} and SSIM~\cite{ssim} metrics that approximate human perception for image similarity comparison. 
Third, we measure how well the restored image preserves the facial {\em identity} of the original image using ID Distance (IDD)~\cite{wang2023restoreformer++}. Unlike the other metrics (that consider features of the {\em entire} images), IDD focuses more on facial features related to a person's identity (e.g., eyes and nose). IDD calculates the angular distance between the feature vectors of the restored face image and the ground-truth image (features extracted by a pre-trained ArcFace~\cite{deng2019arcface}) 

Finally, to assess the overall quality of the restored images, we use FID~\cite{fid} which measures the distribution similarity between the output datasets (of the restored images) and the ground-truth datasets. FID measures how well the restored faces resemble general human faces.

\para{Comparison Baselines.}
We compare \name{} with existing \new{restoration methods including Fantômas~\cite{todt2024fantomas}, GFP-GAN~\cite{gfpgan}, DifFace~\cite{difFace}, and PULSE~\cite{pulse}. 
Fantômas, GFP-GAN, DifFace use auto-encoder, GAN, and diffusion models, respectively, for face restoration. PULSE is a super-resolution model to restore face images. We chose these models because they are able to handle Gaussian blur, and their authors have made the code available for sharing. Fantômas was originally trained with kernel size $K$=29 and thus we retrained it with $K$=37 and 81. We also included a Parrot Recognition method~\cite{parrot_recog} as an identity retrieval baseline. As described in \S\ref{sec:privacy-lit}, Parrot Recognition directly trains a reidentification classifier on {\em blurred} images without performing face restoration. 
} 





\subsection{Basic Evaluation Results}
\label{sec:BasicEvalResults}

\begin{table}
    \centering
    \small
    \begin{tabular}{c|cc} \toprule 
         Kernel Size&  K = 37&  K = 81 \\ \midrule
         Blurred& 38.1\% & 0.0\% \\ \midrule \midrule
         Parrot Recognition& 60.8\% & 51.5\% \\
         Fantômas& 35.0\% & 8.0\% \\
         GFP-GAN& 26.8\% & 1.0\% \\ 
         PULSE& 0.0\% & 0.0\% \\ 
         DifFace& 37.1\% & 1.0\% \\
         Ours& \textbf{100\%} & \textbf{95.9\%} \\ \bottomrule
    \end{tabular}
    \caption{Identity retrieval accuracy. We test both the light-blur ($K=37$) and heavy-blur ($K=81$) settings. 
    }
    \label{tab:IR-Acc}
    \vspace{-0.12in}
\end{table}

We first run basic experiments to evaluate identity retrieval and face restoration under the closed-world setting (i.e., the target identity is included in the reference database).


\para{Identity Retrieval Accuracy.}
We start by evaluating the identity retrieval model (Table~\ref{tab:IR-Acc}). Before testing any face restoration methods, we first establish a baseline by directly running {\em our} identity retrieval method on the {\em blurred images} $x$ without restoration. The result is shown in the ``blurred'' row in Table~\ref{tab:IR-Acc}. The result confirms that ``light blur'' (with $K$=37) is not enough to fully hide the person's identity as the model can still recognize 38.1\% of the {\em blurred faces}. However, under ''heavy blur'' (with $K$=81), none of these blurred faces are recognizable (0\% accuracy). \new{Table~\ref{tab:IR-Acc} also shows the performance of Parrot Recognition~\cite{parrot_recog} where the classifier (ArcFace) has been retrained on blurred images (with matching kernels). While parrot recognition has an improved accuracy of 60.8\% and 51.5\% under the heavy and light blur settings, respectively, the accuracies are still not considered high.}

\begin{figure}[t]
  \centering
  \includegraphics[width=\linewidth]{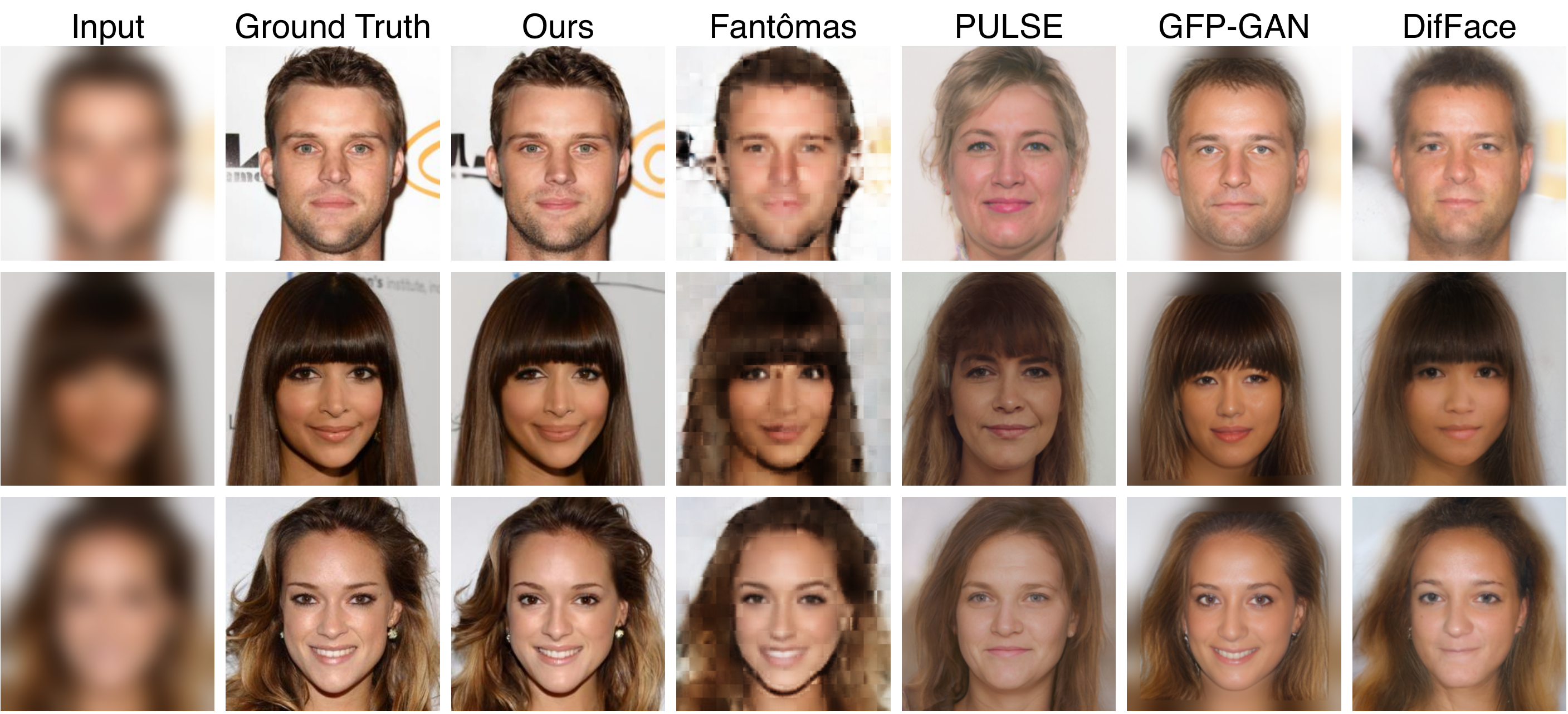}
  \caption{Face restoration under Heavy Blur ($K$ = 81). 
  }
  \label{fig:kernel_81}
\end{figure}

Next, we examine identity retrieval accuracy on {\em restored} faces. For \name{}, we run the identity retrieval model ($M_R$) on the pre-recovered images (from $M_B$) to identify the person in the image. We test both the heavy-blur ($K$=81) and light-blur ($K$=37) settings. As shown in Table~\ref{tab:IR-Acc}, our system achieves much better performance compared with existing methods. Under  light-blur, our system can 100\% recover the facial identity using the pre-recovered images from $M_B$. Under the heavy-blur setting, our system still achieves a 95.9\% identity retrieval accuracy. 
Among the other baseline methods, faces restored by Fantômas, DiFace, and GFP-GAN preserve some identity information under the light-blur setting. However, under the heavy-blur setting, such identity information is no longer recovered, which leads to a low identity retrieval accuracy (lower than 8.0\%). This is confirmed by the examples in Figures~\ref{fig:example} and~\ref{fig:kernel_81}. 
 
The few errors from our system are caused by the inherent challenge of distinguishing identities with similar facial features and makeup. Recall that our reference database contains 28,000 images and 6,084 identities, which makes face-matching difficult. We present case studies in Appendix~\ref{sec:case_study}. As shown in Figure~\ref{fig:mistakes_in_IR}, the mismatched identity looks very similar to the restored face (as well as the original face).

\begin{figure}[t]
  \centering
  \includegraphics[width=\linewidth]{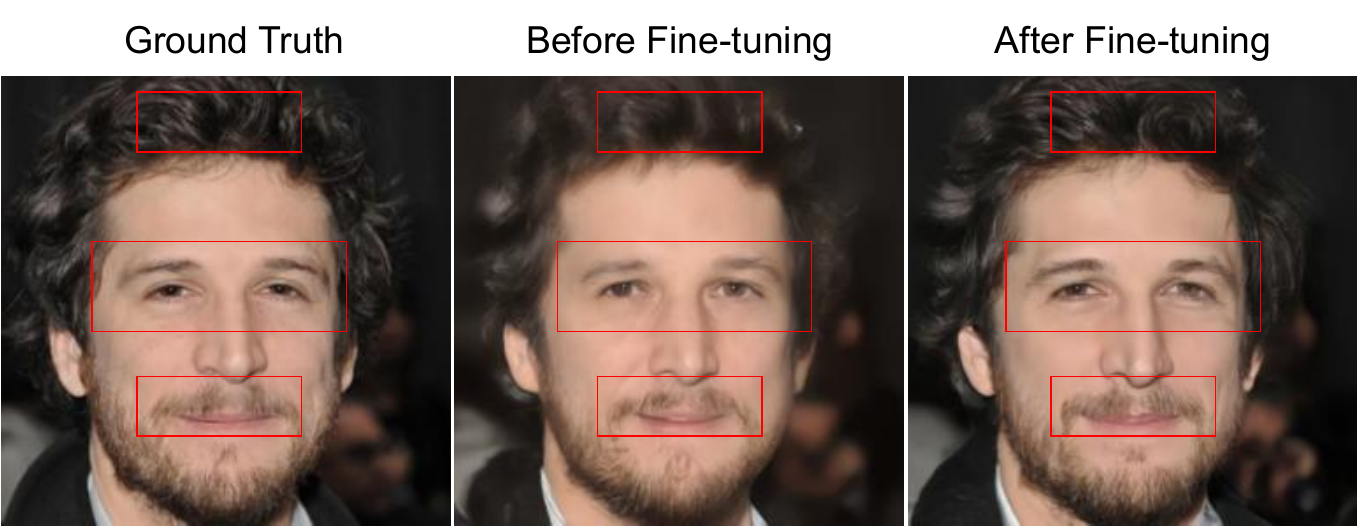}
  \caption{Restored faces {\em before} fine-tuning and {\em after} fine-tuning. We highlight the area where the fine-tuning has made a major improvement in image fidelity and quality. 
  }
  \label{fig:fine-tuning}
  \vspace{-0.1in}
\end{figure}



\begin{table}[t]
    \centering
    \small
        \begin{tabular}{c|ccccc} \toprule 
             Model &  PSNR$\uparrow$ &  SSIM$\uparrow$ &  LPIPS$\downarrow$ & IDD$\downarrow$  & FID$\downarrow$ \\ \midrule 
             $M_B$ & 24.75 & 0.68 & 0.24  & 0.71 & 41.84 \\ \midrule 
             $M_F$ & {\bf 24.94} & {\bf 0.69} & {\bf 0.23}  & {\bf 0.64}  & {\bf 39.97}\\ \bottomrule
        \end{tabular}
    \caption{The Base Model $M_B$ vs. the Fine-tuned Model $M_F$ under the heavy-blur setting ($K = 81$). $\uparrow$ means a higher value is better. $\downarrow$ means a lower value is better.
    }
    \label{tab:fine-tune}
\end{table}

\begin{figure}[t]
  \centering
  \includegraphics[width=\linewidth]{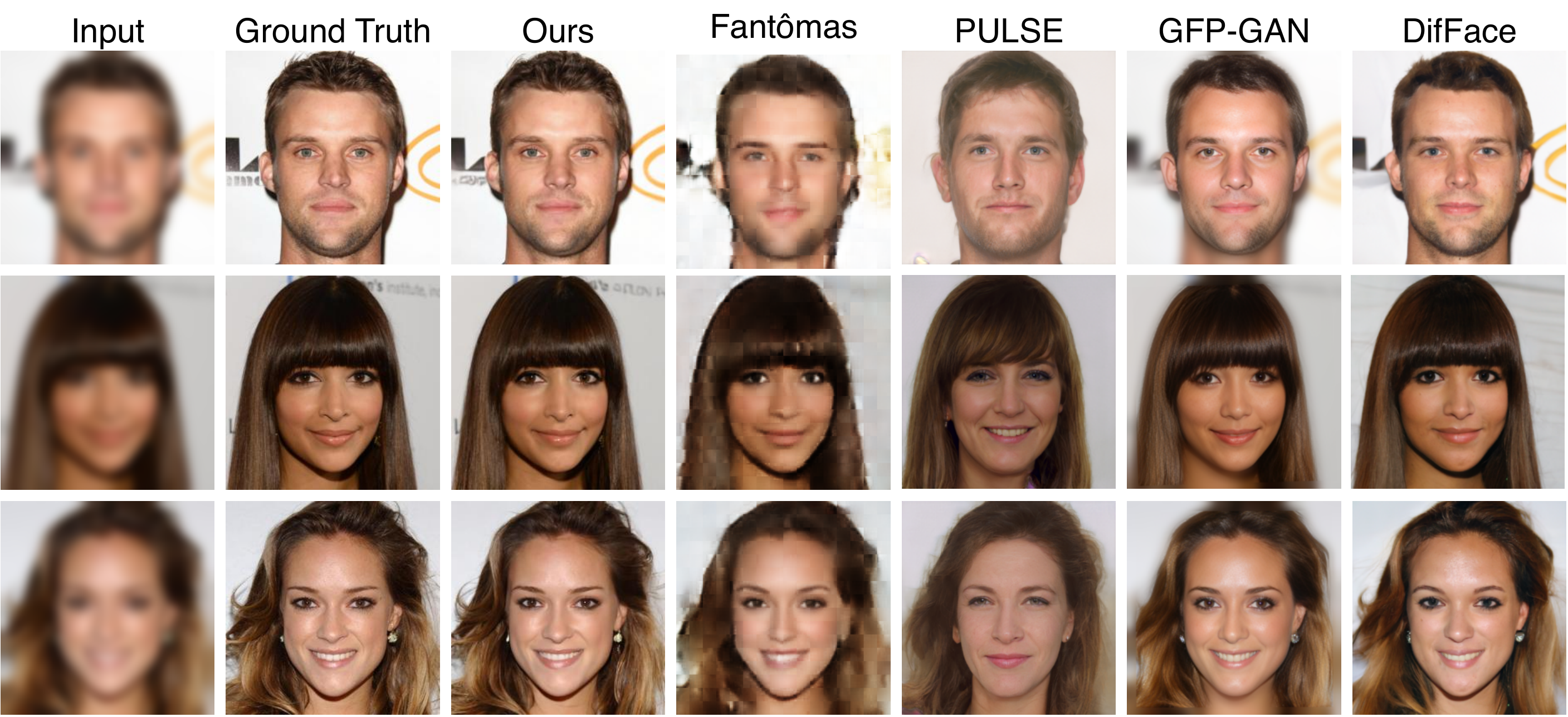}
  \caption{Face restoration under Light Blur ($K$ = 37). 
  }
  \label{fig:kernel_37}
  \vspace{-0.12in}
\end{figure}

\para{Impact of Fine-Tuning.} 
Next, we use experiments to demonstrate the impact of the Fidelity Enhancement module (i.e., fine-tuning) for \name{}. This is done by comparing the restored image quality and fidelity before and after running the fine-tuned model ($M_F$). For brevity, we only run this experiment for the heavy-blur setting ($K=81$) because the image quality is already very good without fine-tuning under light blur (see Figure~\ref{fig:kernel_37}). 

As shown in Table~\ref{tab:fine-tune}, the fine-tuned model $M_F$ outperforms the base model $M_B$ across all evaluation metrics. 
Among the five metrics, we observe noticeable improvements for IDD (which measures the fidelity of the face images with respect to the original {\em identities}).  Through fine-tuning, we can effectively preserve the facial features of the original person and suppress the randomness introduced by the stochastic diffusion process.  

Figure~\ref{fig:fine-tuning} shows a qualitative comparison between the output images of the two models. While the output before fine-tuning (from $M_B$) is already a high-quality face image, its fidelity to the original face is still slightly off. In this example, facial features related to the eyes and the mouth are still different from those of the original person. Also, the hair area still has some blurring effect. After fine-tuning, we can observe that the image (from $M_F$) has clear fidelity improvements in these facial features, making it more similar to the original person. Other subtle differences can be observed by zooming in on the example images.

 \begin{table}
    \centering
    \small
    \begin{tabular}{c|ccccc} \toprule 
         Methods&  PSNR$\uparrow$ &  SSIM$\uparrow$ &  LPIPS$\downarrow$ & IDD$\downarrow$  & FID$\downarrow$  \\ \midrule
         Blurred& 22.87 & 0.66 & 0.61  & 1.32  & 148.19\\ \midrule \midrule
         Fantômas& 23.65 & 0.66 & 0.44 & 0.98 & 120.79\\
         GFP-GAN& 24.08 & 0.69 & 0.29 & 0.90 & 94.80\\ 
         PULSE& 18.78 & 0.54 & 0.43  & 1.37  & 104.72\\ 
         DifFace& 25.30 & 0.70 & 0.29 & 0.88  & 78.07\\
         Ours& \textbf{28.04} & \textbf{0.78} & \textbf{0.17}  & \textbf{0.37} &  \textbf{27.16}\\ \midrule \midrule
         GT& $\infty$ & 1 & 0 & 0 & 0.1\\ \bottomrule
    \end{tabular}
    \caption{Face restoration comparison under Light Blur ($K$ = 37). ``GT'' denotes the ground-truth clear image. 
    }
    \label{tab:e2e-light}
\vspace{-0.12in}
\end{table}

\begin{table}[t]
    \centering
    \small
    \begin{tabular}{c|cccccc} \toprule 
         Methods&  PSNR$\uparrow$ &  SSIM$\uparrow$ &  LPIPS$\downarrow$ & IDD$\downarrow$  & FID$\downarrow$ \\ \midrule 
         Blurred& 19.38 & 0.56 & 0.61  & 1.50  & 290.51\\ \midrule \midrule
         Fantômas & 22.06 & 0.61 & 0.49  & 1.16 & 134.61\\
         GFP-GAN & 19.89 & 0.57 & 0.38  & 1.28 & 123.68\\
         PULSE& 17.25 & 0.50 & 0.46  & 1.42  & 116.78\\
         DifFace& 20.49 & 0.57 & 0.38  & 1.26  & 99.94\\
         Ours& \textbf{24.94} & \textbf{0.69} & \textbf{0.23}  & \textbf{0.64} & \textbf{39.97}  \\ \midrule \midrule
         GT& $\infty$ & 1 & 0 & 0 & 0.1\\ \bottomrule
    \end{tabular}
    \caption{Face restoration comparison under Heavy Blur ($K$ = 81). ``GT'' denotes the ground-truth clear image. 
    }
    \label{tab:e2e-heavy}
    \vspace{-0.12in}
\end{table}






\para{Comparison with Existing Methods.}
Finally, we compare the performance of \name{} with existing methods. 
Under the light-blur setting $(K = 37)$, we compare the results from our model (without fine-tuning) with the existing models. As shown in Table~\ref{tab:e2e-light}, our model (even without fine-tuning) outperforms all baseline methods across all evaluation metrics. The most noticeable improvements are shown by the LPIPS, FID, and IDD metrics. These metrics measure perceptual similarity, overall image quality, and identity-level fidelity. We present example images in Figure~\ref{fig:kernel_37}. Under the light-blur setting, existing methods demonstrate their ability to restore blurred images. The overall image quality is good even though the restored faces may not always look like the original person.




\name's advantage is more visible under the {\em heavy-blur} setting $(K = 81)$. We compare the results of existing models with our $M_F$ model. As shown in Table~\ref{tab:e2e-heavy}, our model outperforms existing models on all evaluation metrics, with a much bigger gap than that under the light-blur setting. Figure~\ref{fig:example} has already shown some example images, and we present additional examples in Figure~\ref{fig:kernel_81} to cross-compare with the examples under light blur (Figure~\ref{fig:kernel_37}). 
On one hand, existing methods are not identity-aware. As a result, they cannot faithfully restore the identity of the original person in the image. In contrast, our method can achieve identity-aware face restoration, by combining a conditional diffusion model with identity retrieval. On the other hand, existing methods are not designed to recover faces from heavy blurring. Their performance is worse under such settings. Overall, the result confirms that \name{} can handle severely blurred images that the state-of-the-art face restoration models are unable to (or not designed to) address.

\subsection{Impact of Mismatched Kernel Size}
\label{sec:ks-exp}


So far, our experiments have assumed that the adversary knows the kernel size $K$ of the Gaussian blur. In practice, the adversary will need to infer this information based on the blurred input. In \S\ref{sec:fr}, we described the methodology for kernel size estimation. Here, we evaluate its effectiveness.   

\para{Estimating Kernel Size Based on Blurred Images.}
To train the kernel size estimator model, we randomly sample 1,000 images from the CelebA-HQ's {\em training set}, and split the data randomly into training, validation, and testing sets with 800, 100, and 100 images, respectively. Then, we apply Gaussian blur with different kernel sizes on each image. We set kernel size $K$ by enumerating all the odd numbers from 1 to 81 to generate blurred images. As mentioned in \S\ref{sec:gaussian}, the kernel size needs to be an odd number to align the center pixel. We train the regression model using the training set, adjust hyperparameters using the validation set, and test the model accuracy on the testing set.


The result shows that adversaries can {\em accurately} predict the kernel size based on the blurred image. The mean absolute error (MAE) is only 0.934, meaning that the model can predict the kernel size $K$ with an offset lower than 1. For example, if the ground-truth kernel size is 37, most predictions either fall on the correct kernel size of 37, or fall on the two neighboring kernel sizes (35 or 39).
The result is further visualized in Figure~\ref{fig:kseconfmatrix} in Appendix~\ref{sec:kse-app}. This means adversaries can estimate the kernel size based on input images, and select the right model (trained with the matching kernel size) for image restoration.


\para{Tolerance of Mismatched Kernel Sizes.}
Next, we further examine how much \name{} can tolerate mismatched kernel sizes. For this experiment, we select the heavy-blur setting (i.e., the more challenging setting) and train the Base Model $M_B$ with Gaussian blur using a kernel size of 81. Then for the closed-world testing images, we apply Gaussian blur with varied kernel sizes ranging from 71 to 91. We show that the model can easily tolerate kernel size mismatches within the mean absolute error (MAE) of 6, under which the restored images are still of high quality and fidelity. Visual examples and quantitative metrics are presented in Figure~\ref{fig:transferability_examples} and Figure~\ref{fig:transferability_linegraph} in Appendix~\ref{sec:kse-app}. 
Given the above kernel size estimation model achieves an MAE below 1, we argue that the proposed method can sufficiently tolerate this level of mismatch.  

In addition, we further test a different dependency function between $K$ and $\sigma$ for Gaussian blur. As mentioned in Section~\ref{sec:gaussian}, most Gaussian blur implementations have a fixed dependency function between these two parameters, and thus attackers only need to predict $K$. If future implementations change the dependency function, does \name{} still work? In Appendix~\ref{sec:kse-app}, we have tested different dependency functions between $\sigma$ and $K$ and have an interesting observation. That is, regardless of the $K-\sigma$ dependency, as long as the blurring effect is similar to what \name{} is trained on, the system still works. This means adversaries can use the kernel estimator to blindly predict a kernel size $K$ that has a similar blurring effect, and then select the corresponding $M_B$ for restoration. For example, we use a different dependency function ($K$=27 and $\sigma=6$) to blur the image. Our kernel estimator predicts $K = 39$ (meaning, the blurring effect is similar to  $K=39$ under the {\em old dependency function}). Then we choose $M_B$ trained for the ``light-blur'' setting (with a close kernel size of 37) and it works well. Appendix~\ref{sec:kse-app} also includes additional transferability evaluation for a {\em non-squared kernel} for Gaussian blur and has a similar conclusion.

\subsection{Open-world Setting Experiment}
\label{sec:open-world}




In this section, we evaluate the open-world setting where the input blurred image $x$ has an OOD identity. In other words, the reference database $D$ does not contain images of the person pictured in $x$. \new{As discussed in \S\ref{sec:threat-model}, in this case, the attacker can still attempt to restore the victim's face. In practice, the deblurred photos can still lead to ad-hoc deanonymization, e.g.,  by viewers who know the victim in real life, such as families, friends, and colleagues. 
}

We collected OOD identities and their images in two ways. (1) From the {\em testing set} of CelebA-HQ, we randomly sample images whose identities only appear in the testing set. In other words, these identities do not have any images in the training set (57 images of 51 identities). (2) We take another public face image dataset FFHQ~\cite{karras2019stylebasedgeneratorarchitecturegenerative} and randomly sample 40 identities (and their 40 images). We manually verified that these identities never appear in the reference database. In total, we have 97 images from 91 OOD identities. The OOD dataset is of the same size as the in-distribution set used by the closed-world experiment. 

Our experiment has two goals. First, we test if we can {\em detect} OOD identities from in-distribution identities. 
Second and {\em more importantly}, even when an identity is OOD, we examine if \name{} can still {\em restore} the blurred face.

\begin{figure}[t]
  \centering
  \includegraphics[width=\linewidth]{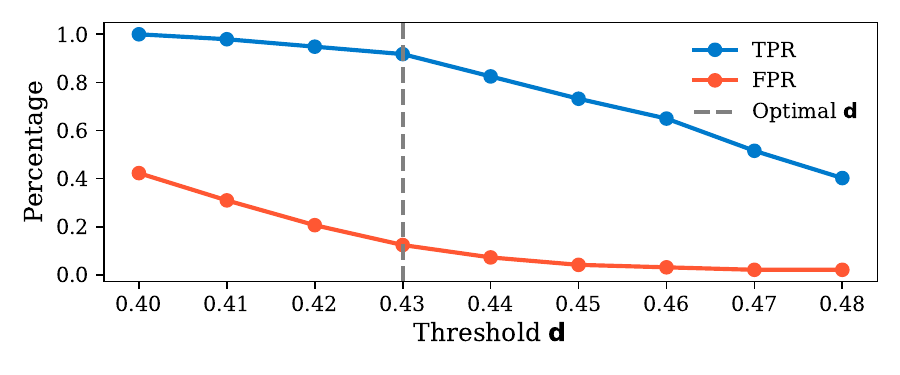}
  \caption{The true positive rate (TPR) and false positive rate (FPR) under different thresholds $d$. The gray line denotes $d=0.43$ which yields an overall accuracy of 90\%.
  }
  \label{fig:rejection_threshold}
\end{figure}

\para{Detecting OOD Identities.}
As discussed in \S\ref{sec:ir}, we apply a threshold $d$ on the lowest distance ($ld_x$) between the pre-recovered image and the reference images, to determine if the blurred image $x$ has an OOD identity. Using the OOD and In-distribution datasets described above, we perform a classification of OOD images under the heavy-blur setting ($K=81$) given it is more challenging for attackers. 

Figure~\ref{fig:rejection_threshold} shows the trade-off between the true positive rate (TPR) and the false positive rate (FPR) when the threshold $d$ varies. 
When we set $d=0.43$, we obtain a TPR of 0.92, an FPR of 0.12, and an overall detection accuracy of 90\%. The result confirms we can detect OOD identities from the blurred images with a decent accuracy. We emphasize that detecting the OOD identities is only to provide {\em contexts} for the adversaries, that is, this person is likely not indexed in the reference database. The more important task is still to restore the blurred face, despite it being an OOD identity.

\begin{table}[t]
    \centering
    \footnotesize
    \begin{tabular}{l|ccccc} \toprule 
         Setting &  PSNR$\uparrow$ &  SSIM$\uparrow$ &  LPIPS$\downarrow$ & IDD$\downarrow$ & FID$\downarrow$ \\ \midrule
        OOD (Detected) & 24.92 & 0.70 & 0.25 & 0.86 & 51.87\\
        OOD (Missed) & 26.12 & 0.74 & 0.23 & 0.78 & 56.58
        \\ \midrule 
        In-Distribution & 24.75 & 0.68 & 0.24  & 0.71 & 41.84 \\ \bottomrule
    \end{tabular}
    \caption{Face restoration quality comparison between OOD identities (including correctly detected and missed OOD identities) and in-distribution identities ($K=81$).
    }
    \label{tab:face-ood}
    \vspace{-0.12in}
\end{table}

\begin{figure}[t]
  \centering
  \includegraphics[width=\linewidth]{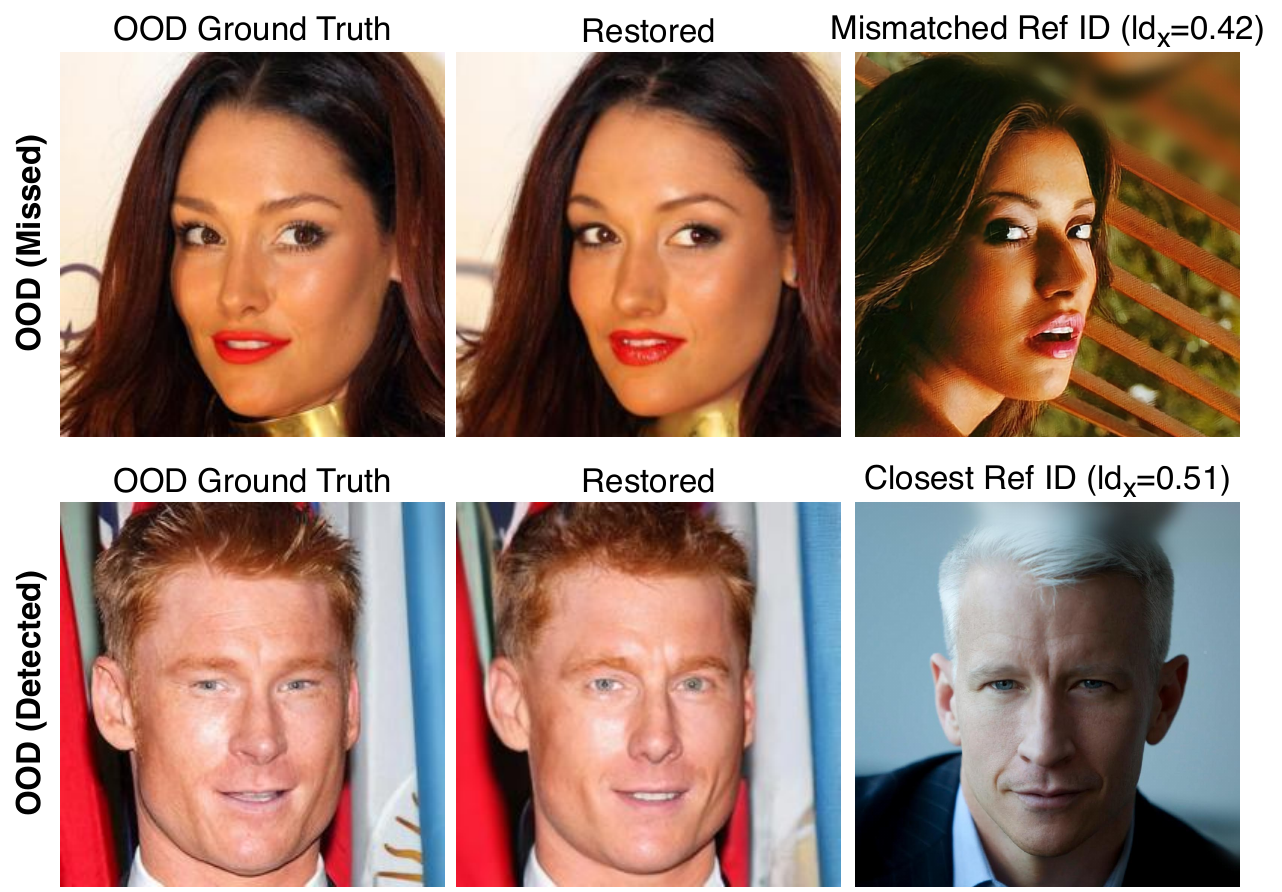}
  \caption{
   Face restoration examples for out-of-distribution (OOD) identities. For both examples, face restoration is successful as the restored face (middle) looks similar to the ground-truth image (left).  
   The top row shows an OOD image that we failed to detect because the reference identity (right) looks too similar to the restored image (middle). The bottom row shows a correctly detected OOD image because the closest reference identity looks different enough. 
  }
  \label{fig:ood-example}
  \vspace{-0.12in}
\end{figure}

\para{Face Restoration for OOD Identities.}
Given OOD identities are not indexed by the reference database, we examine the face restoration result from the Base Model ($M_B$). Table~\ref{tab:face-ood} reports the image quality and fidelity metrics for OOD identities including those that are correctly detected by the detector (using threshold $d=0.43$) and those that are missed. We also show the restoration results for ``in-distribution'' identities as a reference. 
 
We find that the face restoration is highly successful. As shown in Table~\ref{tab:face-ood}, the restored image quality and fidelity are comparable to those of the in-distribution identities. This observation applies to both correctly detected OOD identities and those that are missed. 

Figure~\ref{fig:ood-example} also shows that the restored faces (middle column) look highly similar to the ground-truth images before blurring (left column). We present extra examples in Figure~\ref{fig:OOD_examples-extra} in the Appendix~\ref{sec:extra-imgs}. This result shows  \name{} is able to recover the blurred face even if the corresponding person is never indexed in the reference database. A possible explanation is that \name{} in part learns to approximate the inverse of Gaussian blur. In addition, it may have used other ``similar-looking'' identities to synthesize the facial features for the target image. Figure~\ref{fig:ood-example} shows the closest reference images (right column) in the reference database for both examples. Although the OOD identity does not exist in the reference database, there exist ``similar-looking'' faces in the reference database that can help with face restoration. On one hand, such similar-looking faces are the main reason for OOD detection errors (the top-row example in Figure~\ref{fig:ood-example}). On the other hand, this supports our intuition as to why face restoration is still possible on {\em previously unseen} identities.

\section{Defense}
\label{sec:defense}




Finally, we briefly explore possible countermeasures against the deblurring algorithm and explore adaptive attacks. We want to emphasize that robust defense is not the main focus of this paper. We want to explore a few ideas to inspire future directions. 

The goal of the experiment is to examine how sensitive the deblurring algorithm is to different post-processing steps and mismatched blurring algorithms. More specifically, after Gaussian blur is applied to the image, users can additionally process the blurred image by performing image rotation, adding Gaussian noises, and performing JPEG compression. The user may also use a different blurring algorithm as well. This will create a mismatch between the blurred image and the trained model of the attacker.  
Our experiment seeks to reveal what types of countermeasures have a major impact on image restoration quality.
We run the defense experiments under the light-blur setting ($K=37$), which is a more challenging setting for {\em defenders}.

\begin{figure}[t]
  \centering
  \includegraphics[width=\linewidth]{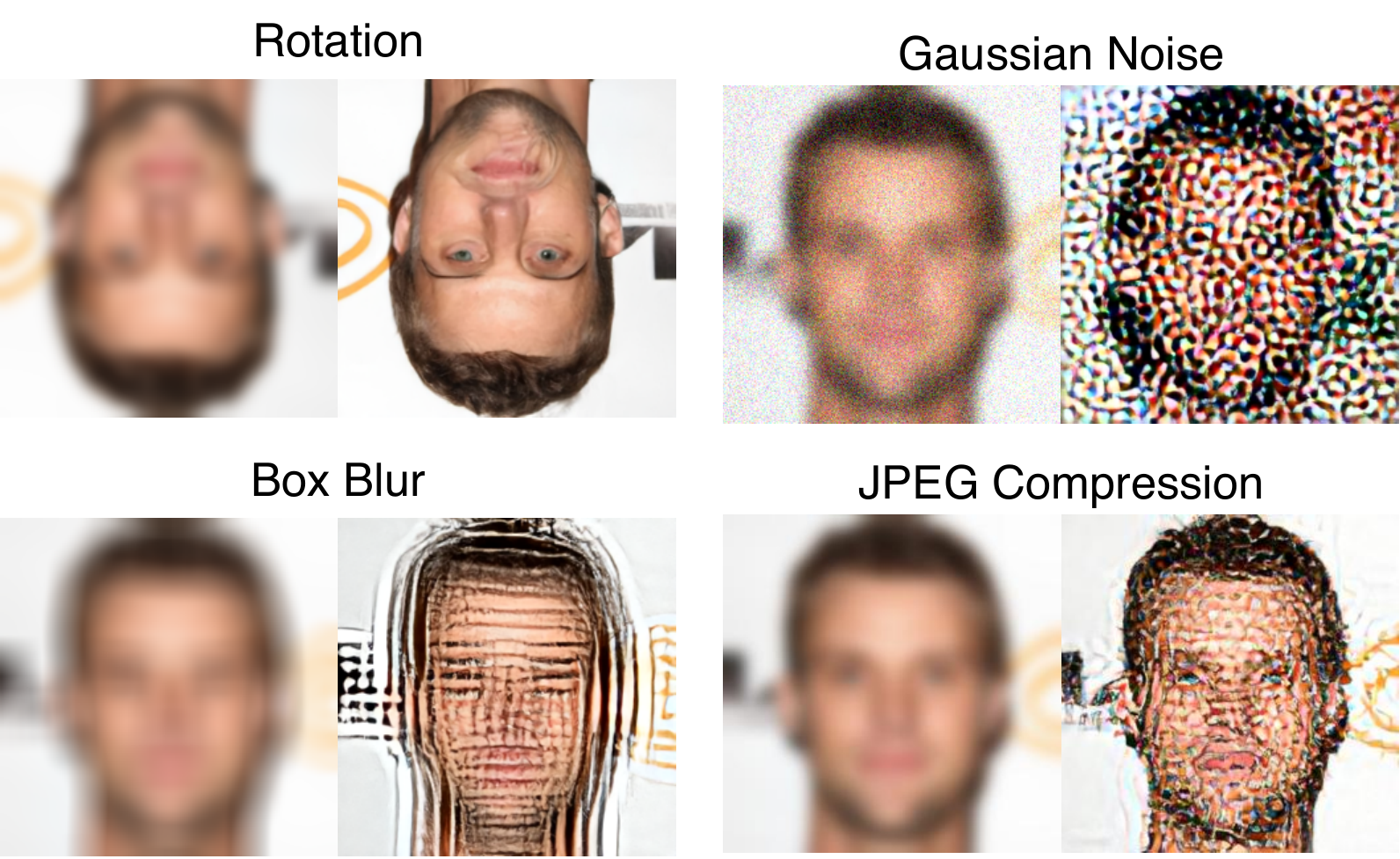}
  \caption{Defense against face restoration. 
  We show that the face restoration method is robust against image rotation, but can be disrupted by other defense methods.
  }
  \label{fig:defense}
\end{figure}

\para{Disrupting Face Restoration Model.}
To disrupt the face restoration model, we apply image rotation, Gaussian noise, and JPEG compression on the blurred image $x$. We also test Box Blur (the OpenCV implementation)~\cite{opencv_library}, a different blurring algorithm, to examine its disruption effect. Appendix~\ref{sec:defense_config} includes further details for the specifications of these defense methods. 
The restoration result is shown in Table~\ref{tab:defense}. We also present visual examples in Figure~\ref{fig:defense}. 
We show that the face restoration is robust against image rotation, but can be majorly disrupted by Gaussian noises, JPEG compression, and box blur. The restored image quality is significantly lower after these post-processing steps (especially for Gaussian noise). In Table~\ref{tab:defense} (the last column), we also report the identity retrieval accuracy (IRA) using the restored faces. For rotation, the IRA is still high (87.62\%). The other defense methods can drop the IRA to 0\%, confirming their effectiveness.

\begin{table}[t]
    \centering
    \footnotesize
    \begin{tabular}{r|cccccc} \toprule 
         Defense&  PSNR$\uparrow$ &  SSIM$\uparrow$ &  LPIPS$\downarrow$ & FID$\downarrow$ & IRA$\uparrow$ \\ \midrule 
         Rotation& 27.39 & 0.76 & 0.23 & 51.69 & 87.62\%\\
         Gaussian Noise & {\bf 11.30} & {\bf 0.07} & {\bf 0.74} & {\bf 425.54} & {\bf 0\%}\\
         JPEG Compress. & 17.15 & 0.25 & 0.63 & 376.37 & {\bf 0\%}\\ 
         Box Blur& 14.62 & 0.23 & 0.59 & 420.76 & {\bf 0\%}\\
         \midrule
         No Defense & 28.04 & 0.78 & 0.17 & 27.16 & 100\% \\ \bottomrule
    \end{tabular}
    \caption{Comparison of different defense approaches (against $M_B$, $K= 37$). To be consistent with other tables, $\uparrow$ means that a higher value is better for {\em attackers} (i.e., worse for defenders). 
    }
    \label{tab:defense}
\end{table}

\begin{table}[t]
    \centering
    \footnotesize
    \begin{tabular}{r|cccccc} \toprule 
         Attack&  PSNR$\uparrow$ &  SSIM$\uparrow$ &  LPIPS$\downarrow$ & FID$\downarrow$ & IRA$\uparrow$ \\ \midrule  
         \makecell{Adaptive Attack}
         & 25.71 & 0.73 & 0.22 & 25.71 & 88.66\%\\ 
         \bottomrule
    \end{tabular}
    \caption{
    Impact of adaptive attack against JPEG Compression Defense ($K=37$).
    $\uparrow$ means that a higher value is better for {\em attackers}.
    }
    \label{tab:adpative_attack}
    \vspace{-0.12in}
\end{table}

\begin{figure}[t]
  \centering
  \includegraphics[width=0.75\linewidth]{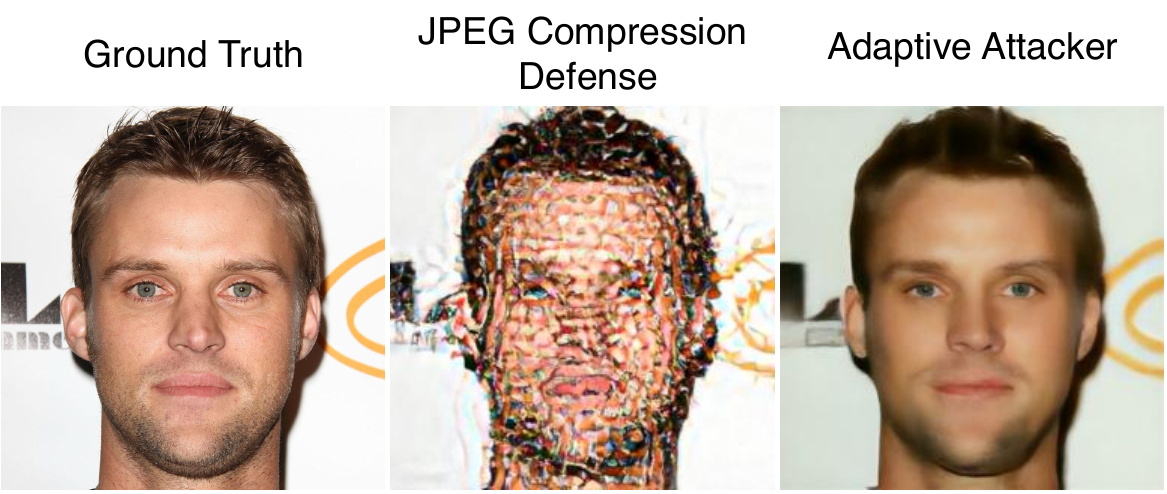}
  \caption{Adaptive attack against JPEG Compression defense. The result shows that the adaptive attack can effectively restore the blurred face. 
  }
  \label{fig:adaptive_attack}
  \vspace{0.1in}
\end{figure}

\para{Adaptive Attacks.}
We want to emphasize that this experiment result {\em does not} mean the defense will be equally effective against adaptive attackers. \new{
To demonstrate this, we implement an {\em adaptive attack} against the JPEG Compression (one of the seemingly effective countermeasures). This is done by fine-tuning our Base Model $M_B$ with images that are processed with Gaussian Blur and JPEG compression (for 100 epochs). The results of the adaptive attack are presented in Table~\ref{tab:adpative_attack}, and example images are provided in Figure~\ref{fig:adaptive_attack}. 
With the {\em adaptive} attack, we observe that the restored face quality is much better. Importantly, the identity retrieval accuracy (IRA) improves from 0\% back to 88.66\%. The result confirms that the adaptive attack is successful. }

As a countermeasure to the adaptive attack, defenders may further randomize the post-processing configurations (including the choice of post-processing methods and their parameters) as a ``secret'', and apply a different blurring process each time for each image. This may make a pre-trained deblurring model less effective on all blurred images. \new{However, we argue that one can also choose {\em not to play this cat-and-mouse game}, for example, by covering the full face using a black/white square. In this way, zero information is left on the photo that is associated with the identity of the target person. 
}

\section{Discussion}
\label{sec:discuss}


\subsection{Implications}
\label{sec:im}

\para{Key Findings.}
Our paper provides concrete evidence that Gaussian blurred face images, even under a {\em high blurring} level, can be restored to their clear form, and can be used to re-identify the persons in the images. We also show a reasonable level of robustness of the deblurring process against unknown kernel sizes. Also, face restoration is possible, not only for {\em known} identities, but also for {\em previously unseen} identities that are not indexed in the reference database.  

\para{Privacy-Utility Tradeoff?}
\new{
Today, Gaussian blur is widely used by lay users and even journalists on sensitive photos in the real world~\cite{faceblur1, faceblur2, faceblur3, faceblur4}. This may be due to the wide availability of Gaussian Blur tools, potentially misleading articles/recommendations on the Internet, and a lack of security awareness among users. When applying Gaussian blur, is there a meaningful trade-off between {\em privacy} and {\em utility}? From the users' perspective, there might be a trade-off since lightly blurred faces can still preserve information such as gender, hair color, and skin tone, making the photo appear more ``authentic''~\cite{todt2024fantomas}. However, the light blur makes face restoration and denonymization easier. } 



\new{
The next question is, would a higher blur level be safe to use?} We can briefly reason the above question from the perspective of {\em conditional diffusion models}. Our model can deblur image $x$ to preserve $x$'s identity by conditioning the denoising process on $x$. As long as $x$ still carries some information from the original image $y$, with sufficient training, it should be feasible to learn the mapping from $x$ to $y$. A higher blurring level may make the training process more expensive, but it should not be impossible. To make Gaussian blur ``safe'', $x$ should be completely independent of the clear image $y$. In this extreme case, for example, $x$ can be a black image with all pixel values set to 0. In other words, we need to cover the {\em full face} with a black mask instead of blurring the face. We want to emphasize that covering the {\em partial face} (e.g., using eye masks) does not work, because the remaining face areas are still potentially re-identifiable~\cite{todt2024fantomas}. \new{Based on this reasoning, we believe Gaussian blur {\em should not} be used when privacy/anonymity is the primary concern. }


\subsection{Recommendations}
\label{sec:rec}

Based on these findings, we make the following recommendations to users, software vendors, and policy makers.

\para{Users.} Users who need to anonymize faces in sensitive photos should not use the Gaussian blur, especially those implemented by popular photo process software (e.g., Photoshop). Such software can be easily studied by attackers to train {\em targeted} deblurring algorithms. This recommendation applies to both lay Internet users and professionals (e.g., journalists) who handle and publish photos of at-risk user populations (e.g., protesters). However, this does not mean other alternative blurring or degradation algorithms (e.g., pixelization, eyemasks, box blur) are secure. These alternative methods are not the focus of this paper, and thus we cannot speak to their security. However, related works have expressed similar concerns about their ability to hide the target information~\cite{cavedon2011getting, hill2016effectiveness}.






\para{Photo-processing Software Vendors.} We recommend photo-processing software adding a {\em warning message} under the Gaussian blur function (and other similar functions). This is to remind or inform users that Gaussian blur is not safe to be used to anonymize faces in sensitive photos. Such a warning message not only protects users from the threat of deanonymization but can also reduce the liability risks of software vendors. The warning should not affect the normal use of the blurring function for non-security/privacy scenarios. 

\para{Policy Makers.} Face anonymization techniques that have known risks of deanonymization (including Gaussian blur) should be discouraged from being used in safety-critical and privacy-sensitive scenarios such as journalism and legal systems. Pushing new standards and policies will require close collaborations between the lawmakers and the technical community.     

\subsection{Ethics Consideration}
\label{sec:ethics}

We are mindful of the ethical implications of our research activities and the results. We believe our research's potential benefits (e.g., discovering vulnerabilities, raising user awareness, improving security practices) outweigh the potential risks. 


\para{Risk vs. Benefit Reasoning.} Our work follows the common practice and the basic principles of {\em offensive security research}~\cite{ethic-vul2010,hacker-2023}. Like many prior works~\cite{wu2024uniid, garofalo2018fishy, sharif2016accessorize, li2014understanding, todt2024fantomas}, our goal is to reveal the security problem in existing solutions and improve the security practice before attackers independently discover and exploit the vulnerability at a large scale against unprepared, vulnerable targets. 
We argue that a ``false sense of security'' is worse than ``known insecurity.'' In our case, without this type of research (including related prior works~\cite{hill2016effectiveness, mcpherson2016defeating, todt2024fantomas}), users may falsely believe that applying a Gaussian blur to their photos is sufficient to hide their identities. This false sense of security could lead to oversharing behaviors on the Internet, exposing more sensitive photos of users, and thus increasing the risk. A key benefit of our research is to provide concrete evidence on the security risk of applying Gaussian blur as a privacy protection mechanism and increase the awareness of related users (e.g., Internet users, journalists, activists). In addition, the result can potentially inform software vendors and policymakers to use or promote more robust privacy protection mechanisms.   
Our experiments require using datasets of human face images. We understand that human face data is a sensitive type of data. In our study, we limit ourselves to only using CelebA-HQ~\cite{karras2017progressive} and FFHQ~\cite{karras2019stylebasedgeneratorarchitecturegenerative}, which are two publicly available benchmark datasets used by common machine learning research~\cite{gfpgan, gpen, psfrgan, difFace, saharia2022palette, kawar2022denoising, lin2024diffbir, DiffBFR, wang2023zeroshot, xia2023diffir, zhu2023denoising}.  

\para{Responsible Disclosure.} \new{We reached out to related parties, including OpenCV, PyTorch, Adobe Photoshop, Apple (iOS SDK team), and FTC (Federal Trade Commission) to disclose our findings and share our recommendations. So far, we have received acknowledgements from Apple, PyTorch, and Adobe, with corresponding teams looking into the issues. We will document further details of our interactions with these vendors before the paper publication.}

\para{Code/Data Sharing.} We will make our research artifacts (code, datasets) available for sharing with other researchers. However, we do not want malicious parties to use the code to cause harm. As safeguards. we will ask requesters to fill out a short form to explain how they plan to use the code and data. We will also verify the requester's identity and affiliation before sharing.

\subsection{Limitations and Future Work}
\label{sec:limit}

Our paper is limited in several aspects. Here, we discuss open questions and opportunities for future work. 

\para{Even Higher Blurring Levels.}
Our experiments use a high blurring level for Gaussian blur ($K$=81 for 256$\times$256 images). This blurring level already renders a 0\% accuracy for facial recognition (see Table~\ref{tab:IR-Acc}). \new{While we did not test a higher blur level, based on our reasoning analysis in \S\ref{sec:im}, restoring face under a higher blurring level will require more expensive training but should not be impossible. We leave experimentation to future work. }

\para{Security-Aware Face Anonymization.}
Researchers have worked on face/image anonymization techniques with provable privacy guarantees~\cite{li2019anonymousnet,cao2021personalized,wen2023divide}. Considering most of these systems are still research prototypes (i.e., not widely used in commercial products yet), we prioritize the analysis of Gaussian blur in this paper. Future research should further investigate the privacy guarantee of these anonymization methods under practical threat models, especially taking into consideration the emerging generative models and the availability of large image datasets with {\em clear} face images.   


 

\para{Dataset and Evaluation.}
Our experiment is limited by the datasets we use. First, the CelebA-HQ and FFHQ datasets contain high-resolution face images that are mostly front-facing. Future work should further investigate the feasibility of deblurring low-resolution face images or images with side faces. \new{Further, future work can examine the impact of the reference database (e.g., in terms of its size and image quality) on the attack effectiveness.}    
Second, we select CelebA-HQ because it contains identity labels on a large number of diverse face images. However, the number of images per identity is low (5 on average), and the distribution is highly skewed. This creates a challenge to the identity-retrieval model because there are not enough reference images for stable/reliable identity-matching. The identity retrieval process can be further improved with a dataset that contains more images per identity.   

\section{Conclusion}

In this paper, we developed a system called \name{} and used a conditional diffusion model to restore Gaussian blurred face images. With extensive experiments, we showed 
that Gaussian blurred faces, even under a high blurring level, can be restored
to their clear form and used to perform accurate re-identification. 
We showed that \name{} can handle input images blurred with an unknown kernel size and the face restoration can be applied to both known identities and previously unseen identities.
Based on our findings, we explored preliminary countermeasures and provided recommendations to users, software vendors, and policy makers.

\begin{small}
\bibliographystyle{IEEEtran}
\bibliography{references}
\end{small}

\begin{appendices}

\section{Case Study: Identity Retrieval Errors}
\label{sec:case_study}

\begin{figure}[t]
  \centering
  \includegraphics[width=0.83\linewidth]{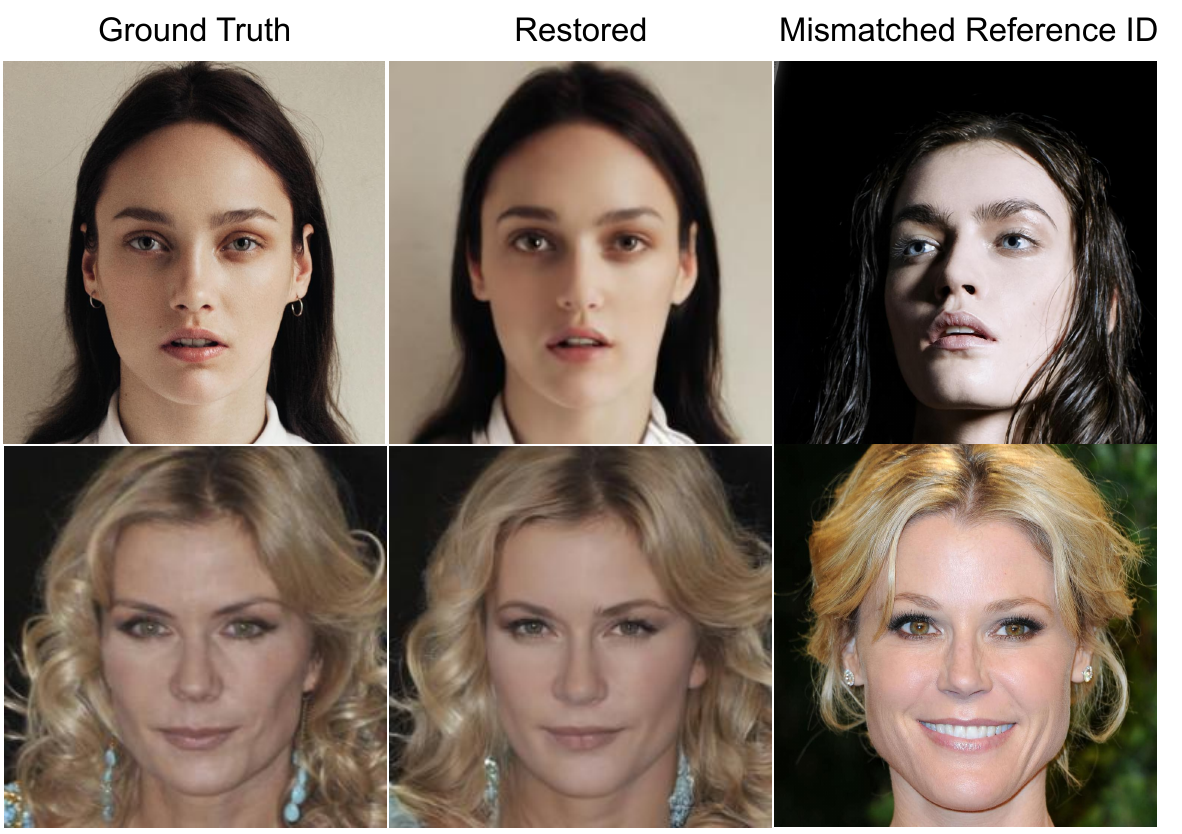}
  \caption{
  Examples of identity retrieval errors. The restored image (middle) from $M_B$ is mismatched with a wrong identity in the reference database (right). In these examples, the restored faces are of high quality. Even though they are matched to a wrong identity (based on CelebA-HQ's identity labels), the mismatched faces also look similar. 
  }
  \label{fig:mistakes_in_IR}
\end{figure}

\begin{figure*}[tb]
  \centering
  \includegraphics[width=0.85\linewidth]{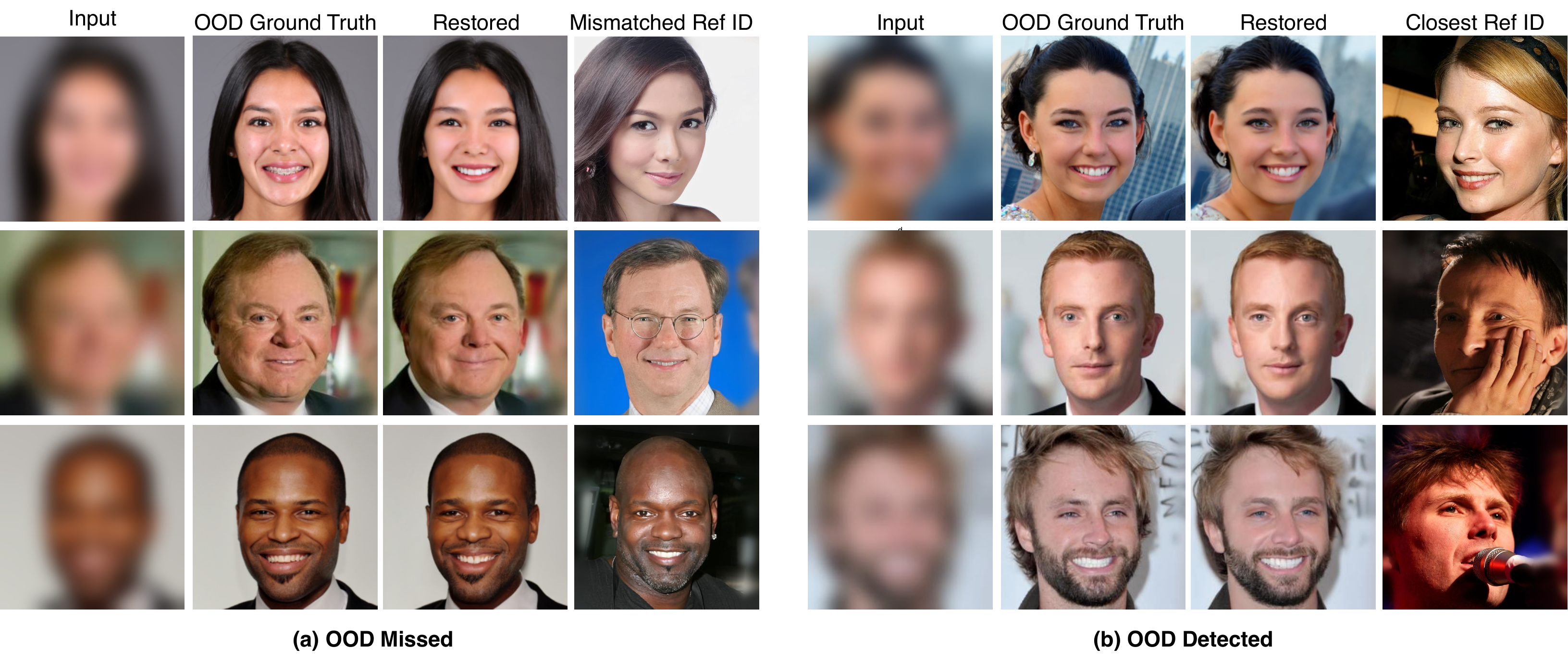}
  \caption{
  Face restoration examples for out-of-distribution (OOD) identities. (a) shows OOD identities missed by our detection method and (b) shows those detected correctly. For all of the examples, face restoration has been successful as the restored faces look similar to the original ground truth.}
  \label{fig:OOD_examples-extra}
\end{figure*}

\begin{figure}[t]
  \centering
  \includegraphics[width=0.8\linewidth]{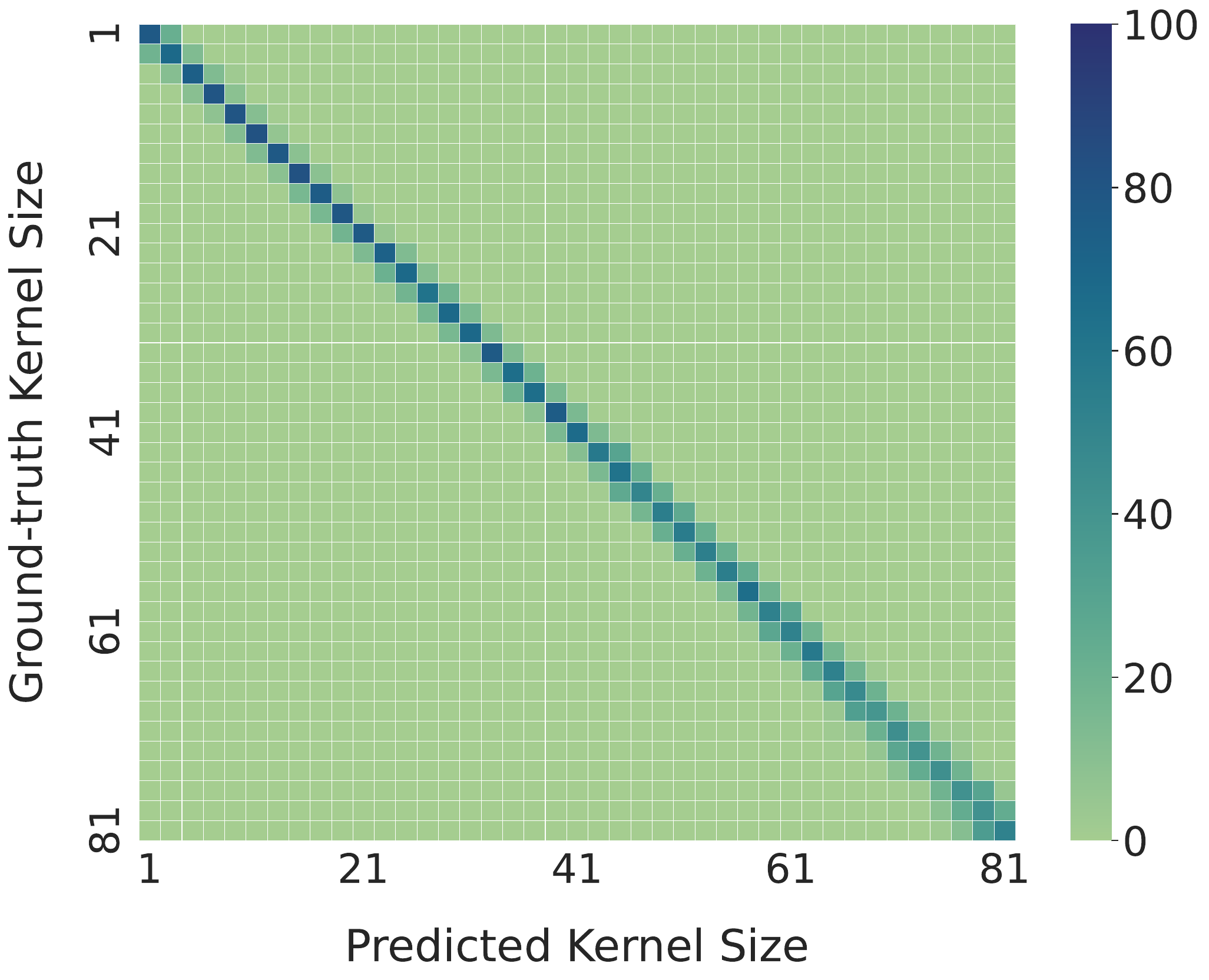}
  \caption{The confusion matrix of the kernel size estimator. The dark squares are aligned with the diagonal line, indicating that the model is highly accurate in predicting the kernel size of the input blurred images. The mean absolute error (MAE) is only 0.934. 
  }
  \label{fig:kseconfmatrix}
\end{figure}

\begin{figure*}[t]
  \centering
  \includegraphics[width=\linewidth]{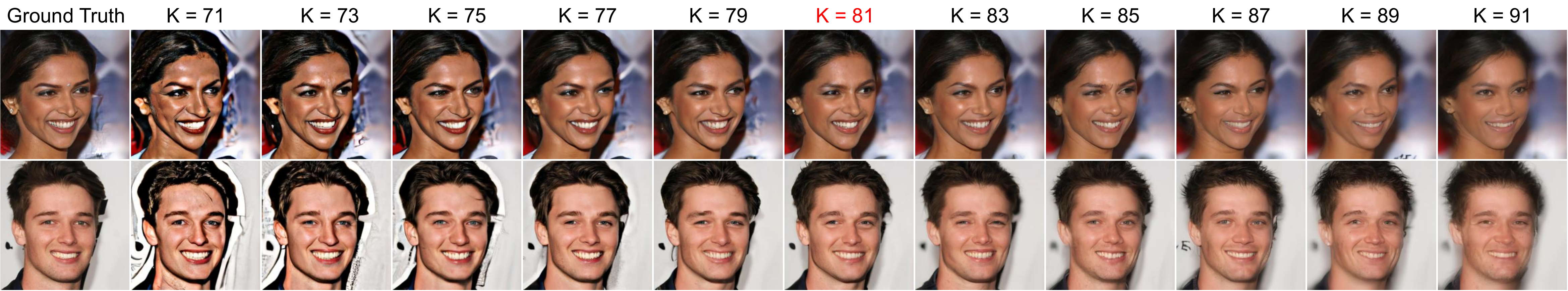}
  \caption{Example face restoration results with mismatched kernel sizes for Gaussian blur. The model is trained with $K = 81$. The testing image is blurred using a kernel size $K$ varying from 71 to 91.
  }
  \label{fig:transferability_examples}
\end{figure*}

\begin{figure*}[t]
  \centering
  \includegraphics[width=\linewidth]{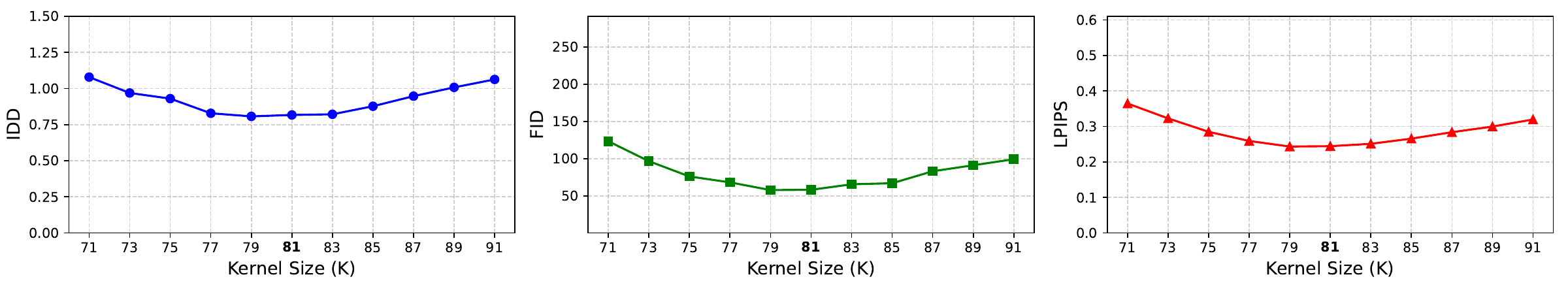}
  \caption{Image restoration quality and fidelity under mismatched kernel sizes. The ground-truth kernel size is $K=81$. Combining with Figure~\ref{fig:transferability_examples}, we show the model can tolerate mismatched kernel size with an offset of 6.  
  }
  \label{fig:transferability_linegraph}
\end{figure*}


\begin{figure}[t]
  \centering
  \includegraphics[width=\linewidth]{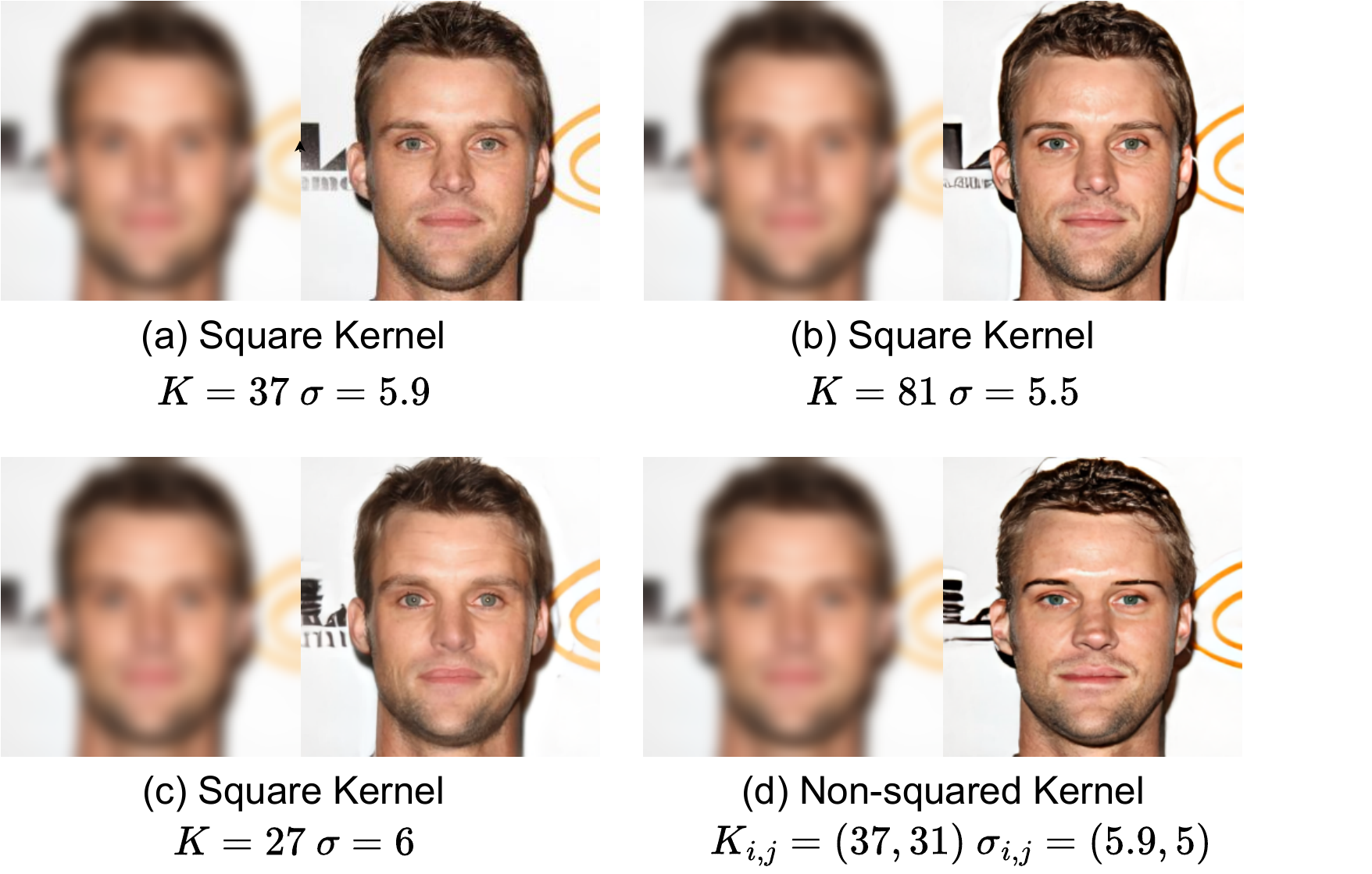}
  \caption{Examples of the transferability experiment results. Each subfigure shows a blurred image (left) and the restored image by our model (right). (a) shows the setting where we use the default $K$-$\sigma$ dependency function in the existing Gaussian blur implementation; (b) and (c) show settings where we use a square kernel with different $K$-$\sigma$ dependency functions; (d) shows the setting where we use a non-squared kernel (with a different $K$ and $\sigma$ on the two dimensions). All these settings have a similar blurring effect. The result shows that the restoration can still be successful under these settings.}
  \label{fig:tolerance}
\end{figure}

As discussed in \S\ref{sec:BasicEvalResults}, our method achieves an identity retrieval accuracy of 95.9\% under the heavy-blur setting. Here, we analyze the error cases. We find that most errors are due to the inherent challenges of facial recognition between identities that look similar. Figure~\ref{fig:mistakes_in_IR} presents two example error cases. For each case, we show their ground-truth image before blurring (left), the restored image (middle), and the matched reference images from the reference database $D$ (right). We observe that the restored image looks reasonably similar to the ground-truth one. However, the restored image is incorrectly matched to a wrong identity in the reference database. This is because one of the face images of this identity (the right column) looks very similar to the restored image. This is an inherent limitation of face recognition within a large reference database (recall that our reference database contains 28,000 images and 6,084 identities).

\section{Extra Example Images}
\label{sec:extra-imgs}

Extra examples of out-of-distribution (OOD) identities are shown in Figure~\ref{fig:OOD_examples-extra}.


\section{Kernel Size and Transferability}
\label{sec:kse-app}

In this section, we present additional evaluation results for the kernel size estimation model, and the experiments to assess the model transferability against different Gaussian blur configurations.  

\para{Kernel Size Estimator Performance.} Figure~\ref{fig:kseconfmatrix} presents the confusion matrix of the Gaussian kernel size estimator. The x-axis shows the predicted kernel size by the model, and the y-axis shows the ground-truth kernel size used to blur the input image. We can observe the prediction is highly accurate with prediction results aligning with the diagonal line of the matrix. The mean absolute error (MAE) of kernel size estimation is 0.934.    



\para{Transferability: Mismatched Kernel Size.} To understand the impact of the mismatched kernel size on our method, we test our base model (trained on kernel size $K=$81) with images blurred with different kernel sizes. More specifically, we vary the kernel size from 71 to 91 to blur the input images to create the mismatch and then let the base model perform face restoration on these images. 
Due to computing resource limitations, we only run one round of face restoration per image for 50 sampled images ($n=1$). Figure~\ref{fig:transferability_examples} shows example images restored by our model. We also present the quantitative metrics to assess the image quality and fidelity in Figure~\ref{fig:transferability_linegraph}. In Figure~\ref{fig:transferability_linegraph}, the y-range is set based on the metric values of the ground-truth images and those of the blurred images, which represent the lower and upper bounds, respectively. The result shows that our model has some transferability over images blurred with mismatched kernel sizes. The restored faces still have a high-level resemblance compared with ground-truth images with a kernel size offset of 6. Recall that our kernel size estimator has an MAE lower than 1, which means that this level of mismatch is not a concern.

\para{Transferability: Different $K-\sigma$ Dependency Functions.}
As mentioned in Section~\ref{sec:gaussian}, most Gaussian blur implementations have a fixed dependency function between the kernel size $K$ and the standard deviation $\sigma$, and thus attackers only need to predict $K$. We explore whether the model still works if future implementations change this dependency function. During these experiments, we have an interesting observation: regardless of how the $K-\sigma$ dependency changes, {\em as long as the blurring effect is similar to what \name{} is trained on, the system still works}. In practice, this means adversaries can use the kernel estimator to blindly predict a kernel size $K$, and select the corresponding $M_B$ for face restoration for images blurred by an unknown $K-\sigma$ dependency function. 

Figure~\ref{fig:tolerance} (a) shows the result from the default $K-\sigma$ dependency function in the existing Gaussian blur implementation ($K=37$). Then in Figures~\ref{fig:tolerance} (b) and (c), we present the transferability experiments where we use different $K-\sigma$ dependency functions. We pick these settings because their blurring effect is similar to $K = 37$ under the old function (which is used to train $M_B$). This is determined by the trained kernel size estimator. The result shows that the restoration can still be successful under these settings. Taking Figure~\ref{fig:tolerance} (c) for example, we use $K=27$ and $\sigma=6$ to blur the image. Our kernel estimator predicts $K = 39$. This means our kernel estimator believes the blurring effect is similar to  $K=39$ {\em under the old function}. Then we choose $M_B$ under the ``light-blur'' setting (which has been trained with $K=37$, close to the predicted kernel size). We find the face restoration still works. The same observations also apply to Figure~\ref{fig:tolerance} (b).



\para{Transferability: Non-Squared Kernel.}
Finally, we test the model transferability to a non-squared kernel. Here we use a kernel of a rectangle shape with different sizes and Gaussian distributions for the two dimensions. We have the same observation: as long as the blurring effect is similar to what \name{} is trained on, the system still works. We can still use the kernel size estimator (trained by the old square kernel) to predict the kernel size and select the corresponding $M_B$ for face restoration. Figure~\ref{fig:tolerance} (d) demonstrates an example. The image is blurred by a non-squared kernel with $K = (37, 31)$ and $\sigma = (5.9, 5)$. The kernel estimator shows the blurring effect is similar to $K=37$ under the old square kernel. In this case, the face restoration is still successful.

\section{Defense Configurations}
\label{sec:defense_config}
The configurations for the defense methods used in our experiments are as follows: 
(1) Rotation: we rotate each testing image by 180 degrees. (2) Gaussian Noise: the mean of Gaussian Noise is 0 and $\sigma$ is 0.1. (3) JPEG Compression: according to the Pillow documents, when PIL images are saved to JPG or JPEG files, the image quality downgrades by 75\% due to the JPEG compression algorithm. We use this configuration for our experiments. (4) Box Blur: the kernel size of the box blur is $31\times31$. This roughly matches the blurring effect of Gaussian blur with $K = 37$. 


\end{appendices}

\end{document}